\documentclass[12pt,preprint]{aastex}
\usepackage{graphicx}
\usepackage{lscape}

\newcommand{\xvec}{{\bf x}}

\newcommand{\vvec}{{\bf v}}
\newcommand{\uvec}{{\bf u}}
\newcommand{\bvec}{{\bf B}}

\newcommand{\lan}{\langle}
\newcommand{\ran}{\rangle}
\newcommand{\be}{\begin{equation}}
\newcommand{\ee}{\end{equation}}
\newcommand{\bea}{\begin{eqnarray}}
\newcommand{\eea}{\end{eqnarray}}
\newcommand{\beax}{\begin{eqnarray*}}
\newcommand{\eeax}{\end{eqnarray*}}
\newcommand{\ba}{\begin{array}}
\newcommand{\ea}{\end{array}}
\newcommand{\bed}{\begin{description}}
\newcommand{\ed}{\end{description}}
\newcommand{\blc}{\begin{list}{$\circ$}{}}
\newcommand{\blb}{\begin{list}{$\bullet$}{}}
\newcommand{\el}{\end{list}}
\newcommand{\ben}{\begin{enumerate}}
\newcommand{\een}{\end{enumerate}}

\newcommand{\hatr}{\hat {\bf r}}

\def\laprox{\mathrel{\hbox{\rlap{\hbox{\lower4pt\hbox{$\sim$}}}\hbox{$<$}}}}
\def\gaprox{\mathrel{\hbox{\rlap{\hbox{\lower4pt\hbox{$\sim$}}}\hbox{$>$}}}}

\begin{document} 


\title{What is the relationship between photospheric flow fields
and solar flares?}

\author{Brian T. Welsch}
\affil{Space Sciences Laboratory, University of California, 
Berkeley, CA 94720-7450; welsch@ssl.berkeley.edu}

\author{Yan Li}
\affil{Space Sciences Laboratory, University of California, 
Berkeley, CA 94720-7450}

\author{Peter W. Schuck} 
\affil{Heliophysics Science Division, Space Weather Laboratory, Code 674,
NASA Goddard Space Flight Center, 8801 Greenbelt Rd., Greenbelt, MD 20771}

\author{George H. Fisher}
\affil{Space Sciences Laboratory, University of California, 
Berkeley, CA 94720-7450}

\begin{abstract}
We estimated photospheric velocities by separately applying the
Fourier Local Correlation Tracking (FLCT) and Differential Affine
Velocity Estimator (DAVE) methods to 2708 co-registered pairs of
SOHO/MDI magnetograms, with nominal 96-minute cadence and $\sim 2''$
pixels, from 46 active regions (ARs) from 1996-1998 over the time
interval $\tau_{45}$ when each AR was within 45$^\circ$ of disk
center.
For each magnetogram pair, we computed the reprojected, average
estimated radial magnetic field, $\tilde{B}_R$; and each tracking method
produced an independently estimated flow field, $\uvec$.
We then quantitatively characterized these magnetic and flow fields by
computing several extensive and intensive properties of each;
extensive properties scale with AR size, while intensive properties do
not depend directly on AR size. 
Intensive flow properties included moments of speeds, horizontal
divergences, and radial curls; extensive flow properties included sums
of these properties over each AR, and a crude proxy for the ideal
Poynting flux, $S_R = \sum |\uvec| \tilde{B}_R^2$.
Several quantities derived from $\tilde{B}_R$ were also computed,
including: $\Phi$, total unsigned flux; $R$, a measure of the 
unsigned flux near strong-field polarity inversion lines (SPILs);
and $\sum \tilde{B}_R^2$. 
Next, using correlation and discriminant analysis, we investigated the
associations between these properties and flares from the GOES flare
catalog, when averaged over both $\tau_{45}$ and shorter time windows,
of 6 and 24 hours.
Our AR sample included both flaring and flare-quiet ARs; the latter
did not flare above GOES C1.0 level during $\tau_{45}$. 
Among magnetic properties, we found $R$ to be most strongly associated
with flare flux. 
Among extensive flow properties, the proxy Poynting flux, $S_R$, was
most strongly associated with flare flux, at a level comparable to
that of $R$.
All intensive flow properties studied were more poorly associated with
flare flux than these extensive properties.
Past flare activity was also associated with future flare occurrence.
The largest coefficients of determination from correlations with flare
flux that we performed are $\sim 0.25$, implying no single variable
that we considered can explain variations in average flare flux.
\end{abstract}


\section{Introduction}
\label{sec:intro}

\footnotesize 
(A version of this paper with much higher-resolution
images is available at \\
http://solarmuri.ssl.berkeley.edu/$\sim$welsch/public/manuscripts/welsch\_etal\_20090904.pdf)
\normalsize

Solar flares are intermittent releases of energy from the solar
corona, in the form of accelerated electrons and ions, and strongly
enhanced electromagnetic radiation over a wide range of wavelengths,
from radio to X-ray.  Consideration of the gravitational, thermal, and
magnetic energy densities in the corona suggests that only the coronal
magnetic field contains sufficient energy to account for the estimated
energy release on flare time scales \citep{Forbes2000}.  Observations
of stable, non-potential (e.g., twisted and/or sheared) magnetic field
structures associated with coronal energy releases (e.g., sigmoids
associated with coronal mass ejections [CMEs] studied by
\citealt{Canfield1999}), have led to wide acceptance of the storage
and release model of solar flares (see, e.g.,
\citealt{Linker2001}). In this model, magnetic free energy --- energy
above that of the potential magnetic field matching the same
normal-field boundary condition at the base of the corona --- is built
up and stored in the corona on relatively long time scales (from
several hours to weeks) and then suddenly released on much shorter
time scales (minutes to a few hours). This naturally poses two
questions: (1) What processes lead to the buildup of magnetic free
energy in the corona?  (2) And what processes lead to the sudden
release of free magnetic energy in flares?

Free magnetic energy in the corona can increase by: (a)
flows that carry new flux from beneath the photosphere into the
corona, and/or (b) photospheric flows that twist, shear, and/or cancel
\citep{Welsch2006} already-emerged fields. Observations and
theoretical models have demonstrated the viability of these
mechanisms. \citet{Leka1996} convincingly demonstrated that twisted
(non-potential) fields emerge through the photosphere, carrying free
magnetic energy with them.  Simulations of this emergence (e.g.,
\citealt{Magara2001,Fan2001}) produced sigmoidal (S- or
inverse-S-shaped) field lines that may become visible in X-rays (e.g.,
\citealt{Fan2003}), and resemble the CME-associated sigmoids studied
by \citet{Canfield1999}. In already-emerged fields, \citet{Brown2003}
found rotating sunspots in some eruptive active regions, and
\citet{Welsch2004} observed shear flows along the eastern neutral line
in NOAA AR 8210 prior to a flare and CME. In addition, rotating,
shearing, and cancelling photospheric flows have been used to inject
free magnetic energy in simulations (e.g., \citealt{Lynch2008,
Roussev2004, Amari2003a, Lionello2002, Tokman2002, Antiochos1999a}).
While case studies of active region flows have been conducted (e.g.,
\citealt{Keil1994,Li2004,Deng2006}), investigations of the
quantitative properties of photospheric flows and their relationships
to flares in many active regions are lacking.

To remedy this shortcoming, we employed tracking techniques to
estimate flows from magnetogram sequences of a few dozen active
regions (ARs) that exhibited varying levels of flaring activity.  We
then investigated correlations between properties of the estimated
flows and the occurrence of solar flares.  We also studied
relationships between flare occurrence and several properties of AR
magnetic fields.  Of particular note, two magnetic field properties
that have been previously related to solar flares can serve as
benchmarks for flow-flare associations: the total unsigned magnetic
flux $\Phi$ of the AR at the photosphere (e.g., \citealt{Leka2007});
and Schrijver's (2007) $R$ value, \nocite{Schrijver2007} a measure of
the unsigned flux near strong-field polarity inversion lines (SPILs).

As an aside, we note that Schrijver's $R$ value is very similar to a
measure developed earlier by Falconer et al. (2003, 2006)
\nocite{Falconer2003,Falconer2006} --- the length of SPILs, which
should be strongly correlated with $R$ by virtue of the strong-field
threshold.  In fact, Barnes et al. (2009) \nocite{Barnes2009} report a
correlation coefficient of 0.95 between these two quantities for their
AR sample.  Falconer et al. (2003, 2006) were among the first to
quantitatively characterize the strong correlation between the
presence of strongly non-potential photospheric magnetic fields and
strong gradients in both radial and line-of-sight (LOS) magnetic
fields.  While the work Falconer et al. (2003, 2006) dealt with CMEs,
and that of \citet{Schrijver2007} dealt with flares, these two
phenomena are clearly related (see, e.g., \citealt{Andrews2003}).  As a
practical matter, we found Schrijver's measure simpler to compute than
that of Falconer et al. (2003, 2006).

We are aware of studies of helicity evolution in active regions which
suggest that the flux of magnetic helicity across the photosphere
might play a role in solar flares (see, e.g., \citealt{LaBonte2007}).
Tests of flow estimation methods using simulated data
\citep{Welsch2007}, however, demonstrated that helicity estimates are
often error-prone.  Accordingly, we plan to investigate flare-helicity
relationships in a future study, in which possible sources of error
can be carefully examined.

In the next section, we describe our data set of magnetogram sequences
from a few dozen ARs, and in \S \ref{sec:methods}, we describe how we
estimated velocities from these sequences.  In \S
\ref{sec:properties}, we describe how we quantitatively characterized
the intensive and extensive properties of the magnetic fields and
estimated flows.  In \S \ref{sec:associations}, we explore
associations of these magnetic and flow properties with flares.
Finally, in \S \ref{sec:conc}, we discuss what we have learned from
our study.

\section{Data} 
\label{sec:data}

We chose to estimate flows from magnetic evolution observed in
sequences of MDI \citep{Scherrer1995} full-disk ($\sim 2''$ pixels),
96-minute nominal cadence, LOS magnetograms.  Compared to available
magnetogram sequences with higher resolution and higher cadence
(including some in which the full magnetic vector is estimated), the
near-ubiquitous coverage of these full-disk LOS magnetograms is a
tremendous advantage for tracking.  After we began this study, the
SOI/MDI team recalibrated field strengths in the MDI full-disk,
96-minute database; as our analysis had already progressed
substantially, we chose to stick with the Level 1.8.0 data, which only
included the plate scale correction.\footnote{See
http://soi.stanford.edu/magnetic/Lev1.8/}

While sequences of white-light data could have been tracked instead,
it is unclear precisely how white-light evolution is related to the
underlying flow field.  Further, evolution of the photospheric
magnetic field is presumably more directly related to evolution of the
coronal magnetic field --- which powers flares --- than evolution of
photospheric white-light intensity.  In addition, the relatively short
lifetime of white-light intensity features --- most notably granules
--- places stringent limits on data cadence.  Since magnetic
structures appear much more persistent than white-light features,
tracking with slower cadences is possible (though not necessarily
desirable, a point discussed further below).

We selected several dozen ARs and AR remnants from MDI synoptic maps
from 1996 to 1998, from solar minimum into the ascending phase of the
most recent cycle.  Our sample was unbiased with respect to flare and
CME productivity, but not with respect to AR morphology: we chose ARs
that were, subjectively, both roughly bipolar and isolated (at the
photosphere). Some of the AR remnants that we selected and tracked
lacked sunspots, and therefore also lacked a NOAA AR designation;
these have been excluded from the results presented here. The AR sample
we use here is comprised of $N_{AR} = 46$ ARs.

Properties of our AR data set are presented in Table
\ref{tab:ar_data}.  We only tracked each AR while it was within about
45$^\circ$ of disk center as it transited the disk; the start and stop
times for tracked intervals are shown in the table.
%
%
To present a quantitative measure of the variability of flare activity
within our sample, we have also included a measure of each AR's
average daily GOES flux from flares at or above the C1.0 level,
\be \bar{\cal F}_{45} = (100S^{(X)} + 10S^{(M)} + 1.0S^{(C)})/ \tau_{45}
~, \label{eqn:avg_power} \ee
where $\tau_{45}$ is the time interval (in days) that the AR was
tracked across the portion of the disk within 45$^\circ$ of disk center,
and $S^{(i)}$ is the sum of GOES flare significands in the $i$-th GOES
class over $\tau_{45}$,
after \citet{Abramenko2005}.  The units of $\bar{\cal F}_{45}$ are
$\mu$ W m$^{-2}$ day$^{-1}$.  To associate flares with ARs in our
sample, we used flare source designations from the NOAA flare
catalog,\footnote{http://www.ngdc.noaa.gov/stp/SOLAR/ftpsolarflares.html\#xray}
although this catalog is known to contain biases
\citep{Wheatland2001}.  Flares associated with our ARs, during the
time interval that each AR was tracked, include: zero A-class flares,
119 B-class flares, 154 C-class flares, 15 M-class flares, and two
X-class flares.  While the true frequency of occurrence of A- and
B-class flares should exceed that of C-class flares, based upon the
distribution of flare peak fluxes (e.g., \citealt{Crosby1993}), weak
flares are often not ascribed to any AR in the GOES catalog, and we
have only included flares with a source AR in our study.  Of our 46
ARs, only 24 flared above the GOES C1.0 level while within 45$^\circ$
of disk center; of the remaining 22, a further 8 were identified as
the source region of at least one sub-C-class flare.

\section{Methods} 
\label{sec:methods}

The MDI instrument measures the LOS field strength, $B_{\rm LOS}$,
averaged over each pixel. To estimate the radial field, $B_R$, from
the observed $B_{\rm LOS}$, cosine corrections were applied to the LOS
field in each pixel, 
\be B_R = B_{\rm LOS}/\cos(\gamma) ~, \ee 
where $\gamma$ is the heliocentric angle from disk center to each
pixel, equal to the inverse sine of the distance (in pixels) from disk
center to each pixel divided by the solar radius (in pixels).

To compensate for foreshortening, triangulation was used to
interpolate the $B_R$ data --- regularly gridded in the plane of the
sky, but irregularly gridded in spherical coordinates $(\theta,\phi)$
on the solar surface --- onto points $(x,y)$ corresponding to a
regularly gridded Mercator projection of the spherical surface.  This
projection was adopted because it is conformal (locally
shape-preserving), which is necessary to ensure displacements measured
by our tracking algorithm were not biased in direction. The price of
conformality is distortion of scale, such that apparent displacements
are exaggerated by a factor of the secant of apparent latitude; but
this is easily corrected after tracking.  Details of our deprojection
procedure are discussed in Appendix \ref{app:merc}.


We separately used the Fourier Local Correlation Tracking (FLCT;
\citealt{Fisher2008}) and the Differential Affine Velocity Estimator
(DAVE; \citealt{Schuck2006}) codes to produce two estimates of flows
for each pair of successive magnetograms in our sequences.  Prior to
tracking, each pair of magnetograms was cross-correlated and then
co-registered to sub-pixel accuracy using cubic-convolution
interpolation.  For FLCT, pixels below a 20 G threshold in
$|\bar{B}_R|$, where $\bar{B}_R$ is the average magnetic field for
each magnetogram pair, were not tracked, nor were pixels farther than
$\gamma = 45^\circ$ from disk center.  DAVE estimated flows for many
pixels that fell outside of these thresholds, but a filter was applied
to the raw DAVE results that zeroed out flows from pixels where FLCT
did not estimate flows.  Because the physical length corresponding to
Mercator pixels shrinks toward the poles, all estimated flows were
corrected for the distortion in magnitude introduced by the Mercator
projection.  Moving cropping windows were applied to the deprojected
magnetograms to extract subimages containing the ARs as they crossed
the disk.

While the magnetogram cadence was nominally $\Delta t = 96$ minutes,
data gaps did exist.  Flows estimated from magnetogram pairs with
$\Delta t > $ six hours apart were excluded from the analyses of flow
properties described in \S \ref{sec:properties}.  With $\Delta t = 96$
minutes, pixels of $\Delta s \sim$ 1400 km, and flows of $\sim 0.3$ km
s$^{-1}$ (cf., \citealt{Hathaway2002}), displacements would be $\sim
1.2$ pixels.  We expect that this cadence effectively averages away
flows with characteristic time- and length-scales shorter than a
couple of hours and a couple of megameters.  This implies that the flows we
measure ought to be long-lived and large-scale relative to granular
flows typically tracked from sequences of intensity images in
white-light (e.g., \citealt{November1988}) or G-band.

\citet{Demoulin2003} asserted that tracking methods applied to
magnetogram sequences estimate an apparent flow $\uvec$, with
\be \uvec \equiv \vvec_h - v_r \bvec_h / B_r
~, \label{eqn:dnb} \ee
and where $\vvec$ is the plasma velocity, $\hatr$ is the normal to the
magnetogram surface, and $h$ denotes components in the magnetogram
surface.  Here, we are using $B_R$ as an estimate of $B_r$, the true
normal (radial) field.  Substitution of equation (\ref{eqn:dnb}) into
the normal component of the magnetic induction equation produces a
continuity-like equation for the normal component of of the magnetic
field and the flux transport velocity $\uvec$
\citep{Demoulin2003,Schuck2005}.  \citet{Demoulin2003} pointed out
that knowledge of $\uvec$ and $\bvec$ is sufficient to compute the
fluxes of magnetic energy and helicity across the magnetogram
surface. However, recent tests of tracking methods on MHD simulation
data by \citet{Welsch2007} cast doubt on the accuracy of methods that
exclusively track the normal component of the magnetic field. A
followup investigation by \citet{Schuck2008} using the same MHD
simulation data suggest that some tracking methods estimate $\vvec_h$
much more accurately than $\uvec$ --- they produce a biased estimate
of the horizontal plasma velocity.  This supports the view that
line-of-sight tracking methods are insensitive to the vertical flows
that drive flux emergence \citep{Demoulin2009}. However, these methods
may produce a reasonable estimate of the fluxes associated with
footpoint twisting in mature active regions.  Tests conducted by both
\citet{Welsch2007} and \citet{Schuck2008} suggest that velocity
estimates that employ the full magnetic vector --- not just the LOS
data that we use here --- would be more accurate.  Currently, however,
sequences of vector magnetograms are too rare for statistical studies
relating magnetic evolution to flares.

For tracking each active region's data cube, both FLCT and the DAVE
require selecting a spatial windowing parameter, $\sigma$.  In
principle, this choice can be made objectively, by selecting the value
of $\sigma$ that maximizes agreement between the observed magnetic
evolution $\Delta B_R/\Delta t$ between magnetogram pairs, and the
expected relationship \citep{Demoulin2003} between this evolution and
the observed flow $\uvec$,
\be \nabla_h \cdot (\uvec \bar{B}_R) = - \Delta B_R/\Delta t
~, \label{eqn:ctty} \ee
where $\bar{B}_R$ is the average of $B_R$ from each magnetogram in the
pair.  (But LCT schemes are not formally consistent with this
continuity-like equation; see Schuck 2005.)  \nocite{Schuck2005} In
practice, the cadence in our study was so low that equation
(\ref{eqn:ctty}) was relatively poorly obeyed (compared to tests with
other data sets with higher cadence) for any choice of $\sigma$.
Tests that quantified agreement of DAVE results with equation
(\ref{eqn:ctty}) using slope-fitting, as well as linear (Pearson) and
rank-order correlations, suggested that $\sigma$ in the range of $8 -
10$ pixels maximized consistency across these measures.  Accordingly,
$\sigma=9$ pixels was used with the DAVE, and $\sigma = 8$ pixels was
used with FLCT.  We expect this windowing leads to further spatial
smoothing of flows, averaging over flows with characteristic length
scales less than several megameters.  Following \citet{Welsch2004},
positive- and negative-flux pixels were tracked separately with FLCT,
with these separate estimates combined into all-polarity flow maps.

The FLCT code used here was slightly modified from that described by
\citet{Fisher2008}.  The modified version\footnote{available at
http://solarmuri.ssl.berkeley.edu/$\sim$fisher/public/software/FLCT/C\_VERSIONS/}~%
removes the mean intensity from the subimages prior to cross
correlating them.  A Gaussian spatial frequency filter was used in the
FLCT runs, with a roll-off wavenumber in $x$ of $k_{x,{\rm roll-off}}
= 0.25 \, {\rm max}(k_x)$, and similarly for $y$.  This has been found
to improve accuracy in tests with known shifts \citep{Fisher2008}, and
to ameliorate FLCT's tendency to underestimate speeds.

\begin{figure}
\plotone{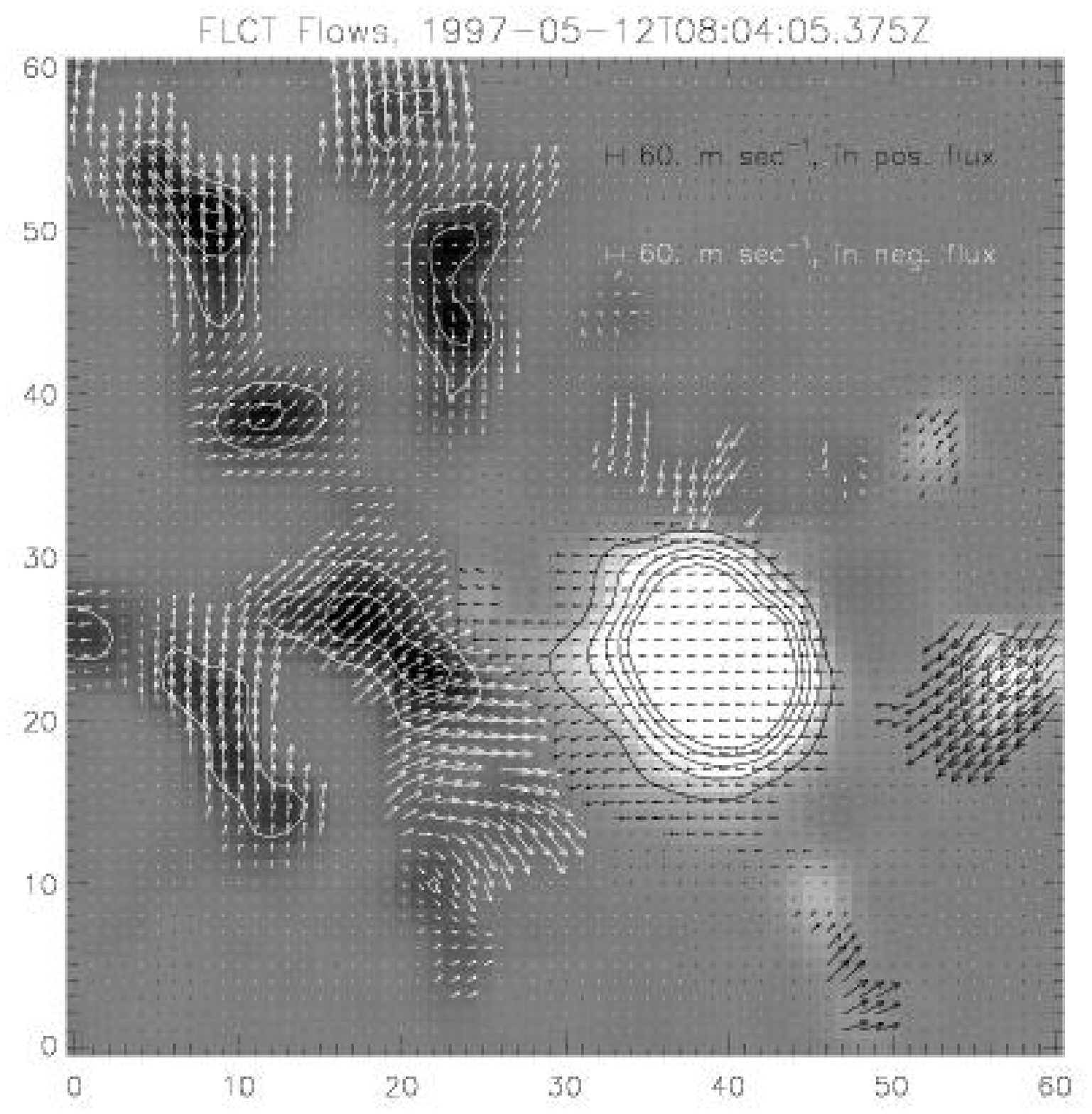}
\caption{%
FLCT flows overlaid on a grayscale image of $\bar{B}_R$ from AR 8038;
white and black vectors represent flows in negative and positive
regions, respectively.
White and black contours of $\bar{B}_R$ are overplotted in negative
and positive regions, respectively, at 100 G intervals between 100 G
and 500 G.
The time shown at top corresponds to the end of the 96-min. interval
over which flows were estimated.  
\label{fig:flct_example}}
\end{figure}

\begin{figure}
\plotone{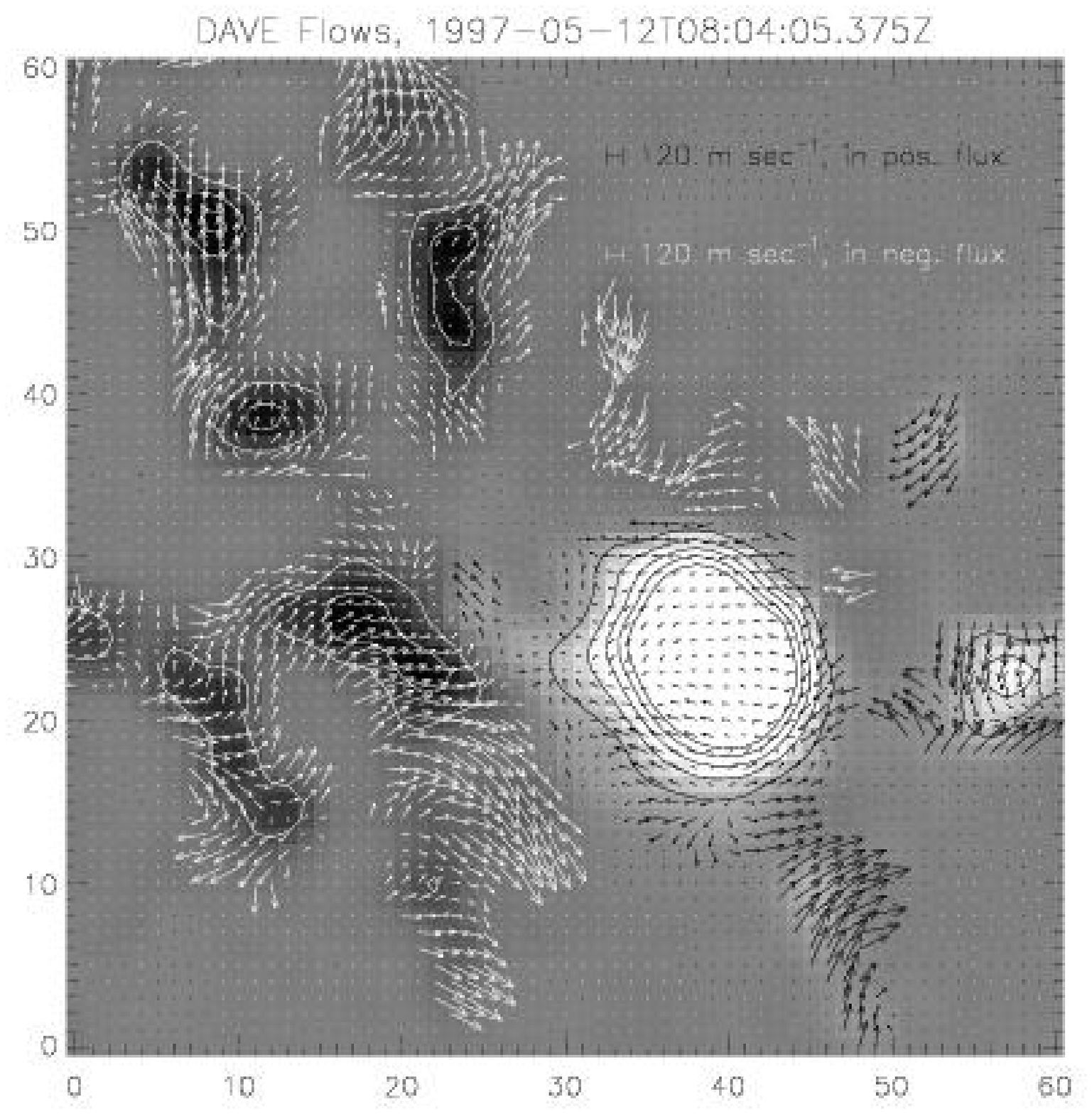}
\caption{%
DAVE flows overlaid on a grayscale image of $\bar{B}_R$ from AR 8038;
white and black vectors represent flows in negative and positive
regions, respectively.
White and black contours of $\bar{B}_R$ are overplotted in negative
and positive regions, respectively, at 100 G intervals between 100 G
and 500 G.
The time shown at top corresponds to the end of the 96-min. interval
over which flows were estimated.  
Note that the vector length scale differs from that in Figure
\ref{fig:flct_example}.
\label{fig:dave_example}}
\end{figure}

\begin{figure}
\plotone{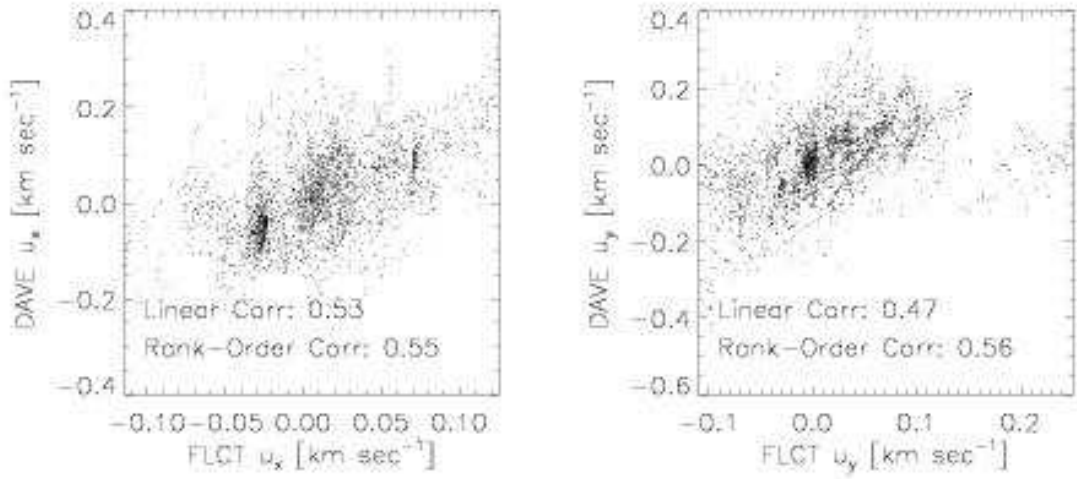}
\caption{%
FLCT and DAVE flows from Figures \ref{fig:flct_example} and
\ref{fig:dave_example}, respectively, are compared in scatter plots of
$u_x$ (left) and $u_y$ (right).  We also show linear and
rank-order correlation coefficients between the flow components.
\label{fig:flct_dave}}
\end{figure}

Our full data set consists of $N_{\rm flow} = 2708$ sets of flows
estimated by both FLCT and DAVE.  A typical FLCT flow field is shown
in Figure \ref{fig:flct_example}.  This example was taken from AR
8038, an active region which produced three flares and CMEs as it
crossed the disk in May 1997 \citep{Li2009}.  The time shown on the
plot corresponds to the end of the 96-minute interval over which the
flows were estimated.  DAVE flows from tracking the same data are
shown in Figure \ref{fig:dave_example}.  While there are broad regions
of agreement between the directions of FLCT and DAVE flows, there are
also some regions where the flows are in the opposite direction.  Our
focus here is not to compare FLCT and DAVE flows, but rather to test
the sensitivity of flow-flare correlations (or lack thereof) to the
method chosen to estimate flows.  Nonetheless, in Figure
\ref{fig:flct_dave}, we show scatter plots relating the $x$- and
$y$-components of FLCT and DAVE flows from the same time interval, to
illustrate how the flows compare in this case.  We only plotted
results for the set of pixels in which both methods estimated flows.
On the same plot, we also show the linear (Pearson) and rank-order
correlations between the flow estimates. Clearly, the methods' flow
estimates are significantly correlated, but do not agree in detail.
For the $\sim 16.5$ million pixels for which both FLCT and DAVE
estimated velocities, the linear and rank-order correlations of the
$x$-components of the flows were 0.67, while for the $y$-components,
these correlations were 0.60 and 0.61, respectively.
%
%
Also, the linear and rank-order correlations between FLCT and DAVE
speeds, when both are weighted by the unsigned field strengths, are
higher still --- 0.85 and 0.76, respectively.  This suggests the flows
tend to agree more closely in strong field regions.

Comparing speeds shows a systematic difference.  Over the same set of
$\sim 16.5$ million pixels, FLCT's mean and median flow speeds were
0.069 and 0.062 km sec$^{-1}$, while for DAVE, mean and median flow
speeds were 0.112 and 0.099 km sec$^{-1}$.  In tests with synthetic
magnetograms from ANMHD simulations, \citet{Welsch2007} found that an
older version of FLCT than that used here underestimated speeds by
about $\sim 30 \%$, which would suggest a DAVE/FLCT slope of $0.7^{-1}
\sim 1.4$.  Here, the DAVE/FLCT speed ratio is nearer $1.6.$

These flow speeds are significantly less than flows estimated by other
methods, e.g., supergranular flow speeds determined from
center-to-limb Doppler variation by \citet{Hathaway2002}.  They are,
however, only slightly slower than flows estimated by
\citet{Chae2004b} for 96-minute cadence data; our slower speeds might
result from our larger spatial windowing, $\sim 17$'' (see their
Fig. 7).

\section{Quantifying Magnetic Field and Flow Properties}
\label{sec:properties}

\subsection{Typical Fields and Changes}
\label{subsec:basic}

To familiarize the reader with our data set, we present some
statistics regarding typical magnetic fields and field changes in the
approximately 21 million pixels in which $|\bar{B}_R|$ exceeded 20 G
(of about 85 million total pixels in our data set).  Over this set of
pixels, the median and average cosine-corrected, unsigned,
pixel-averaged field strengths $|\bar{B}_R|$ were 63 G and 113 G,
respectively.  The median and average unsigned rates of change,
$|\Delta B_R|/\Delta t$ were $3 \times 10^{-3}$ G s$^{-1}$ and $4
\times 10^{-3}$ G s$^{-1}$, respectively, which correspond to unsigned
changes of 17 G and 23 G, respectively, for the nominal $\Delta t =
5760$ s time interval between magnetograms.  Perhaps unsurprisingly,
this is roughly the quoted noise level for MDI
magnetograms.\footnote{See http://soi.stanford.edu/magnetic/Lev1.8/}
The linear and rank-order correlation coefficients between unsigned
$|\bar{B}_R|$ and $|\Delta B_R/\Delta t|$ were 0.36 and 0.41,
respectively --- hence, {\em absolute} changes tend to be larger in
stronger fields.  The linear and rank-order correlation coefficients
for $|\bar{B}_R|$ and $|(\Delta B_R/\Delta t)/B_R|$ were -0.30 and
-0.39, respecively, implying {\em relative} changes tend to be smaller
in stronger fields.  The rank-order correlation between signed
$\bar{B}_R$ and $\Delta B_R/\Delta t$ was 0.05, which amounts to a
small but statistically significant bias: negative fields were
slightly more likely to get more negative, and positive fields were
slightly more likely to get more positive.  This imbalance, however,
cannot be sustained indefinitely. Our ARs all have NOAA designations
(and so possess sunspots), and therefore contain relatively newly
emerged flux; perhaps older flux in decaying active regions
behaves differently.

\subsection{Field Properties} 
\label{subsec:fieldprops}

In this and the following subsections, we describe aspects of magnetic
and flow fields that we calculate to relate to flare activity.  We
distinguish between ``extensive'' properties, which depend directly on
AR size (e.g., integrated quantities), and ``intensive'' properties,
which do not depend directly on AR size (e.g., averaged quantities).
This is in analogy to thermodynamics, in which extensive quantities
scale with the system size (e.g., total thermal energy, or particle
number), while intensive properties do not (e.g., temperature, or
particle density).  As a rule, quantities that would increase if an
AR's length scales were doubled (all else held equal) are extensive,
e.g., total unsigned flux; quantities that would not increase are
intensive, e.g., average magnetic field strength.
%
%
Although our characterization of $\Phi$ as extensive is based upon a
hypothetical doubling of AR area, we note that \citet{Fisher1998}
found that $\Phi$ is in fact strongly positively correlated with (and
nearly a linear function of) AR area.  While \citet{Fisher1998} used
the term ``extrinsic'' for $\Phi$ and other integral field properties,
we prefer extensive for its more direct analogy to thermodynamics.

While our analysis centers on properties of estimated flows, we also
characterized properties of the time-averaged estimated radial
magnetic field $\bar{B}_R$ corresponding to each pair of field
estimates used for tracking, as well as properties of changes in the
magnetic field that do not require tracking.  These magnetic field and
field evolution properties are described below, and then summarized in
Table \ref{tab:props}.

In the subsequent statistical analyses of magnetic field properties,
it will often be appropriate to use magnetic field values scaled in a
way to compensate for the latitude-dependent area distortion
introduced by the Mercator projection.  This rescaling is necessary
because the physical area represented by Mercator-projected pixels
shrinks toward the poles (the effect which causes Greenland and
Antarctica to appear unrealistically large in Mercator projections of
the Earth's surface); details are given in Appendix \ref{app:merc}.
Without rescaling, field strengths in Mercator pixels that represent a
smaller area on the Sun would be given equal weighting in our
analyses.  The apparent area $da$ of a Mercator-projected pixel ---
equal to the MDI disk-center pixel area, $(\sim 1.4$ Mm)$^2$ --- is
related to Mercator-corrected pixel area $d\tilde{a}$ via
\be d\tilde{a}(\tilde{\theta}) = da \cos^2 \tilde{\theta}
~, \label{eqn:rescale} \ee 
where $\tilde{\theta}$ denotes the latitude of each pixel (cf.,
co-latitude $\theta$ often used in spherical coordinates).  Pixel
values used in calculations involving local magnetic properties or
local image structure (e.g., the field strength threshold for
tracking, or contours of $\bar{B}_R$) do not require Mercator
correction.  In contrast, pixel values in global quantities (e.g.,
sums, averages, and higher moments of quantities) must be
weighted by $\cos^2 \tilde{\theta}$.  Because we frequently refer to
Mercator-corrected magnetic field values, we define
\be \tilde{B}_R  \equiv \bar{B}_R \cos^2 \tilde{\theta}
~, \label{eqn:btilde} \ee 
and
\be \tilde{B}_R^2  \equiv \bar{B}_R^2 \cos^2 \tilde{\theta}
~. \label{eqn:b2tilde} \ee 

\citet{Leka2007} demonstrated that the total unsigned flux was
strongly associated with flaring, so we calculated this extensive
quantity from each average estimated radial field,
\be \Phi = \sum |\bar{B}_R| d\tilde{a} = \sum |\tilde{B}_R| da
~, \label{eqn:phi} \ee 
where the sum runs over the cropping window.  

For each magnetogram, we also computed the first four moments of the
estimated radial magnetic field, $\tilde{B}_R$: mean $\lan \tilde{B}_R
\ran$, variance $\sigma^2[\tilde{B}_R]$, skewness $\zeta[\tilde{B}_R]$,
and kurtosis $\kappa[\tilde{B}_R]$, after \citet{Leka2003b}.  We denote this
family of moments ${\cal M}[\tilde{B}_R]$,
\be
{\cal M}[\tilde{B}_R] \equiv \left \{ \begin{array}{ll}
\lan \tilde{B}_R \ran & {\rm mean} \\
\sigma^2[\tilde{B}_R] & {\rm variance} \\
\zeta[\tilde{B}_R] & {\rm skewness} \\
\kappa[\tilde{B}_R] & {\rm kurtosis} ~.
\end{array} \right .
 \label{eqn:cal_m} \ee
Since each of these moments does not, of necessity, scale with active
region size, these descriptors of AR magnetic fields are intensive.

Because flare processes might be nonlinearly related to photospheric
field strength (because, for instance, magnetic energy density is
proportional to magnetic field squared), we also computed
\be  \sum \tilde{B}_R^2  \label{eqn:b2tot} 
~, \ee
where, as above, the sum runs over all pixels in each average
magnetogram.  While $\tilde{B}_R^2$ has units of magnetic energy
density, we do not ascribe any particular physical meaning to the sum
of this quantity; it is included in our analysis to provide baseline
comparisons for other quantities.  (We also investigated sums of
higher powers of $\tilde{B}_R$, but linear and rank-order correlations
of summed $\tilde{B}_R^2$ with summed $|\tilde{B}_R|^3$ and $\tilde{B}_R^4$
are all above 0.9, implying these variables contained little
additional information.)

\citet{Schrijver2007} correlated the total unsigned flux, $R$, near
strong-field PILs (SPILs) with flaring, so we also computed $R$ for
each magnetogram in our data set.  This entails: (1) creating two
separate positive and negative bitmaps, representing regions of strong
average field ($|\bar{B}_R| > 150$ G) in each polarity; (2) computing
the product of dilated versions of these bitmaps, to find loci with
strong, opposite polarity fields in close proximity; (3) computing a
SPIL-weighting function, $W_{SPIL}$, by convolving the product bitmap
with a normalized Gaussian of 15 Mm FWHM; (4) summing the unsigned
flux in the SPIL-weighted magnetogram,
\be R = \sum W_{SPIL} |\tilde{B}_R| 
~.  \label{eqn:rdef} \ee
A few of these steps are illustrated explicitly in
\citet{Welsch2008b}.  Some magnetograms in our study have $R=0$.
While $R$ quantifies one aspect of the internal arrangement of flux
within an AR --- how much lies near strong-field polarity inversion
lines --- it also scales with active region size, albeit with the length
$L$ of each AR, not area $A$.  Accordingly, we also characterize $R$ as
extensive.

In addition to $\Phi$ and $R$, we calculated the time derivatives
$\dot \Phi$ and $\dot R$ between averaged magnetograms (each in [Mx
sec$^{-1}$]), to investigate the correlation between changes in these
quantities and flares.  If the flux emergence rate per unit area, in
Mx sec$^{-1}$ cm$^{-2}$ , were uniform over the Sun, then the nearly
linear dependence of $\Phi$ with the area $A$ over which magnetic flux
is present \citep{Fisher1998} would imply a nearly linear relationship
between $\dot \Phi$ and $\Phi$ --- essentially, the larger an AR's
$\Phi$ is, the larger the cross-section for emergence into that AR. In
fact, \citet{Harvey1993} showed that emergence rates are enhanced over
the background level in areas where flux has already emerged.  Hence,
we characterize $\dot \Phi$ as extensive, and expect $\dot \Phi$ and
$\Phi$ to be correlated.  To account for this, we also calculated
intensive measures of the rates of change of $\Phi$ and $R$, the
flux-normalized $\dot \Phi / \Phi$ and $\dot R / R$, to compensate for
correlations between $\dot \Phi$ and $\Phi$, and $\dot R$ and $R$.

\citet{Wang2006} reported changes in the center-of-flux positions of
each polarity associated with five flares (X- or M- class) in
$\delta$-spot regions.  Accordingly, we also computed the
center-of-flux (COF) positions (Mercator corrected) of each AR
polarity in each average cropped magnetogram,
\be \xvec_\pm = \sum_\pm \xvec \, |\tilde{B}_R| / \sum_\pm |\tilde{B}_R| ~,
\label{eqn:cof} \ee
where the $\pm$ sums run over pixels of the corresponding polarity.
We then computed the mean polarity separation vector,
\be {\bf dx} = \xvec_-  - \xvec_+
~, \label{eqn:dcof} \ee 
which points from $\xvec_+$ to $\xvec_-$, and found the rate of change
$\dot{({\bf dx})}$ (in [km sec$^{-1}$]) of ${\bf dx}$
between pairs of average magnetograms.  We then computed the
``center-of-flux divergence,''
\be D_{COF} = \dot{({\bf dx})} \cdot {\bf dx} /|{\bf dx} |
~, \label{eqn:cofdiv} \ee
and normalized the change by the initial ${\bf dx}$, as well as the
``center-of-flux shear,''
\be S_{COF} = |\hatr \cdot ( \dot{({\bf dx})} 
\times {\bf dx}) | /|{\bf dx} |
~, \label{eqn:cofshear} \ee
implied by these changes. The unit vector $\hatr$ is normal to the
projected magnetogram plane.  Both $D_{COF}$ and $S_{COF}$ have units
of velocity.  A negative (resp., positive) $D_{COF}$ corresponds to
centers of flux growing closer (resp., farther) along the line
separating them.  Based upon flux emergence simulations (e.g.,
\citealt{Abbett2000}), one might expect $D_{COF} > 0$ with the
emergence of new flux, and, conversely, $D_{COF} < 0$ with flux
cancellation.  If centers of flux have a tendency to move closer
together as a result of flares, as \citet{Wang2006} found, then flares
should be anti-correlated with $D_{COF}$.  A non-zero $S_{COF}$
corresponds to displacements of centers of flux in opposite directions
perpendicular to ${\bf dx}$.

\subsection{Flow Properties} 
\label{subsec:flowprops}

\begin{figure}
\includegraphics[scale=0.8]{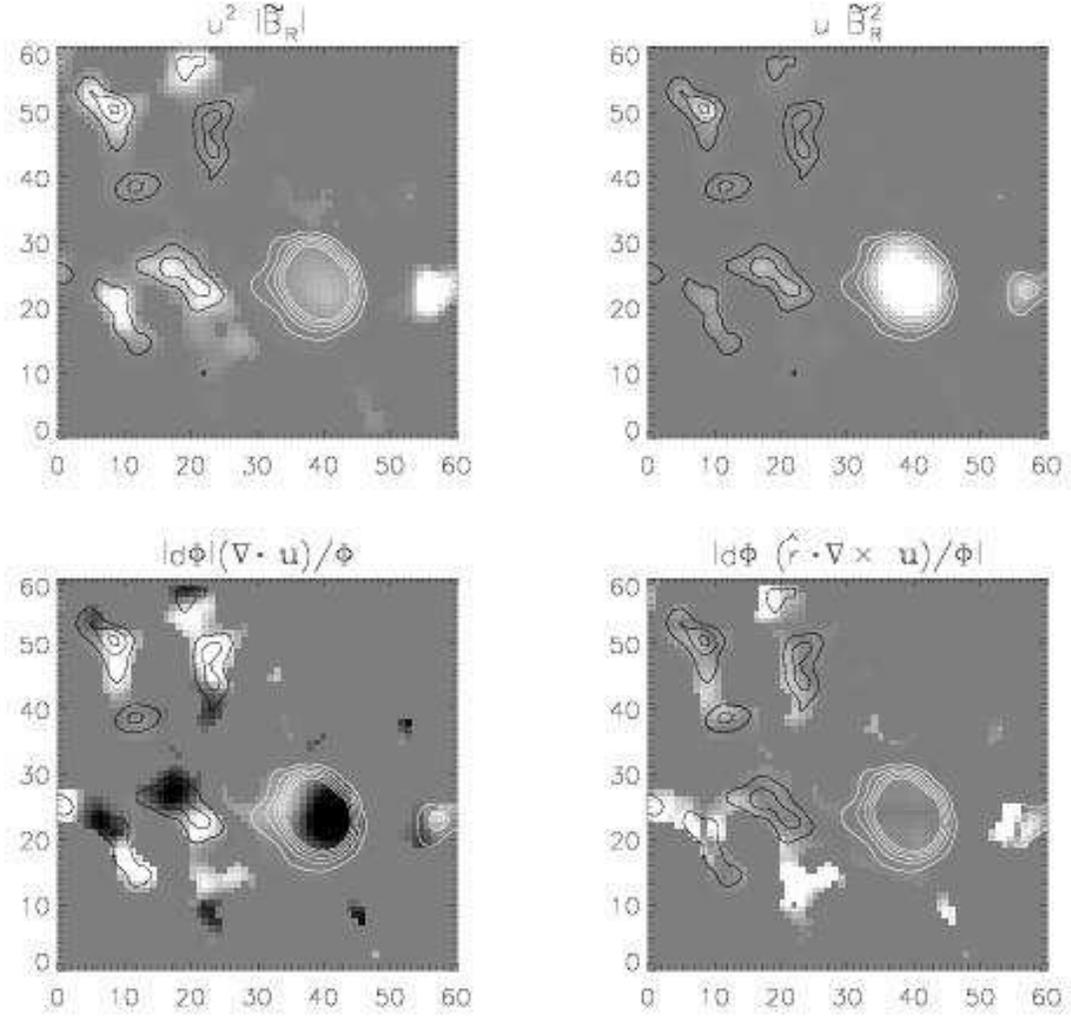}
\caption{%
%
Examples of flow properties from FLCT flows corresponding to the flow
field shown in Figure \ref{fig:flct_example}.  
Top row: Grayscale maps of $u^2 |\bar{B}_R|$ (left) $u \bar{B}_R^2$
(right). Black and white contours of $\bar{B}_R$ are
overplotted in negative and positive regions, respectively, at 100 G
intervals between 100 G and 500 G, for all panels.
Bottom row: Grayscale maps of $|d\Phi|$-weighted versions of the
horizontal divergence (left) and unsigned vertical curl (right).  
\label{fig:maps}}
\end{figure}

Here, we describe how we quantified apects of our estimated flows;
these properties are also summarized in Table \ref{tab:props}.  For
each pair of magnetograms we track, we typically have two independent
flow estimates, by FLCT and DAVE, for each of several thousand pixels.
From these estimates, we computed several intensive measures of flow
properties.  As with $\tilde{B}_R$, we computed the moments of each
flow field, ${\cal M}[u]$.  Because photospheric flows in regions of
stronger $|\bar{B}_R|$ are expected to impact coronal evolution more
strongly than regions with weak $|\bar{B}_R|$, we also computed
normalized and unnormalized flux-weighted moments, ${\cal M}[u
|d\Phi|/\Phi_u]$ and ${\cal M}[u |\tilde{B}_R|]$, respectively.  We
note that $u |\tilde{B}_R$ has units of an electric field, is related
to the horizontal electric field.  In the normalized expression,
$|d\Phi|$ represents the (Mercator-corrected) flux in each pixel, and
$\Phi_u$ is the total unsigned flux only from those pixels for which
each method (FLCT or DAVE) estimated flows. (Averaging over the entire
2D arrays for each velocity estimate --- which contain many zero
velocities [e.g., from weak-field regions] could bias the moments if,
for instance, low-flux ARs have fractionally more weak-field pixels.)
We also computed several related extensive quantities: summed speeds
and field-weighted speeds, $\sum u$ and $\sum u |\tilde{B}_R|$,
respectively; and the sum of a kinetic-energy like quantity, $\sum
u^2$, and its field-weighted version, $\sum u^2 |\tilde{B}_R|$,
respectively.  A grayscale map of $u^2 |\tilde{B}_R|$ is shown in the
upper-left panel of Figure \ref{fig:maps}.

We also computed the sum of a quantity that, dimensionally,
has units of an energy flux,
\be S_R \equiv \sum u \tilde{B}_R^2 
~; \label{eqn:poynting} \ee 
with $u$ in cm sec$^{-1}$ (our default in this paper is km sec$^{-1}$)
and $\tilde{B}_R^2$ in G$^2$, the summand has units ergs cm$^{-2}$
sec$^{-1}$.  Multiplying by MDI's disk-center pixel area $da$, $\sim 2
\times 10^{16}$ cm$^2$, then yields units of power, ergs sec$^{-1}.$
This sum represents an extensive measure of the transport of magnetic
energy {\em at} the photosphere.  The true radial Poynting flux $S_r$
{\em across} the photosphere and into the corona can be derived from
an estimated flow field $\uvec$ and a vector magnetogram
\citep{Demoulin2003} via,
\bea S_r &=& (1/4 \pi) \sum 
[(\bvec_h \cdot \bvec_h) v_r - (\vvec_h \cdot \bvec_h) B_r]
\label{eqn:3comp_poynting} \\
&=& -(1/4 \pi) \sum (\uvec \cdot \bvec_h) B_r 
~, \label{eqn:true_poynting} \eea
which depends sensitively on the relative orientation of the flow and
magnetic fields.  Our simple proxy $S_R$ completely ignores this
dependence.  As long as magnetic fields are not exactly vertical, and
observations reported by \citet{Fisher1998} suggest they tend not to
be, then strong horizontal fields will tend to occur where there are
strong vertical fields.  Further, flux emergence and submergence
occurs on PILs; but our proxy is small along PILs, where the radial
field vanishes.  But flux from emerging (submerging) features should
diverge (converge), which can be detected (see \S
\ref{subsec:flow_expectations}, below).  Consequently, our proxy,
crude though it may be, could be statistically related to actual
energy fluxes.  A grayscale map of $S_R$ is shown in the upper-right
panel of Figure \ref{fig:maps}.

To characterize flow properties arising from the vector nature of our
estimated flows, we computed several additional intensive properties
of the flow.  For each flow field $\uvec$, we computed the absolute
and signed horizontal divergence, and the unsigned radial component of
the vorticity (curl), and then computed moments of these
distributions.  When computing derivatives for these quantities,
Mercator corrections were applied to account for varying pixel scales.
As above, horizontal refers to directions perpendicular to the normal
(radial) direction, $\hatr$.  In addition, we computed the
flux-weighted signed and unsigned horizontal divergences, unsigned
radial curls, and moments of these distributions.  We excluded
derivatives in pixels with any of the four-nearest neighbor pixels
below the absolute field-strength tracking threshold, to avoid
contributions to divergences and curls from flow discontinuities
between tracked and untracked pixels.  Weighted divergence and curl
maps are shown in the left and right (resp.) panels in the bottom row
of Figure \ref{fig:maps}.  In addition, we computed extensive versions
of these quantities, totaling signed and unsigned, unweighted and
field-weighted divergences and curls.

\begin{figure}
\plotone{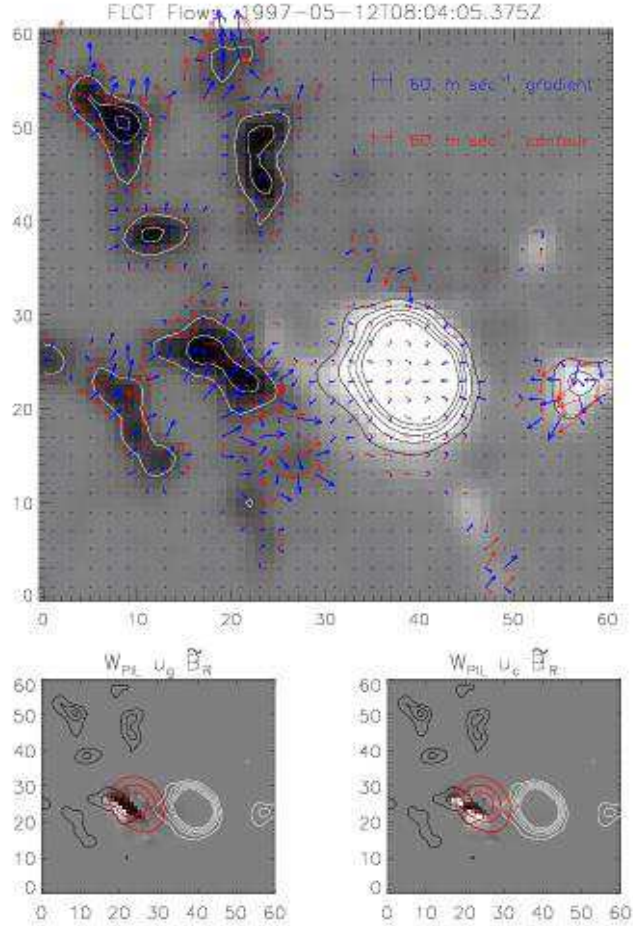}
\vspace{-2.0in}
\caption{%
Top: The background grayscale shows $\bar{B}_R$ (white is positive,
black negative), and white and black contours of $\bar{B}_R$ are
overplotted in negative and positive regions, respectively, at
100 G intervals between 100 G and 500 G.
Components of FLCT flows from Fig. \ref{fig:flct_example} that
lie along gradients and contours of $\bar{B}_R$ are shown with blue
and red vectors, respectively.  
For clarity, not every vector is plotted.
Bottom left: A grayscale map of $W_{PIL}$- and $\bar{B}_R$-weighted gradient
flows, used to quantify converging motions.  
Black and white contours of $\bar{B}_R$ are overplotted in negative
and positive regions, respectively, at 100 G intervals between 100 G
and 500 G.
Red contours of the PIL-weighting map $W_{PIL}$ are overlaid.
Bottom right: A grayscale map of $W_{PIL}$- and $\bar{B}_R$-weighted contour
flows, used to quantify shearing motions.  
All other image components are as with the bottom left image.
\label{fig:gradcont}}
\end{figure}

Numerical simulations of coronal evolution leading to coronal mass
ejections often use shearing flows (e.g., \citealt{Antiochos1999a}) or
converging flows with cancellation (e.g., \citealt{Linker2003}), or
some combination of the two (e.g., \citealt{Roussev2004}), along PILs
to both inject energy into the coronal magnetic field and form
eruptive structures.  In addition, flux emergence is expected to lead
to diverging flows near PILs (see, e.g., \citealt{Abbett2000}).
Basically, these flow patterns can be distinguished by their
relationships with the local gradient in normal field: shearing
velocity vectors point along normal-field contours, and converging or
diverging flows near PILs are parallel or anti-parallel to
normal-field gradients there.  Accordingly, to characterize the
presence of shearing or converging/ diverging flows near PILs in an
automated way, we also decomposed flow estimates $\{ \uvec \}$ into
components along $\xvec_g$, a unit vector that points along the local
horizontal gradient of $\bar{B}_R$,
\be \xvec_g \equiv \nabla_h \bar{B}_R/|\nabla_h \bar{B}_R| 
~ \label{eqn:gradhat} \ee
and along $\xvec_c$, a unit vector that points along contours of
$\bar{B}_R$,
\be \xvec_c \equiv \hatr \times \nabla_h \bar{B}_R/|\nabla_h \bar{B}_R|
~ \label{eqn:conthat} ~. \ee
We denote these (signed) gradient and contour flows as $\{ u_g \}$ and
$\{ u_c \}$, respectively.  Examples of gradient and contour flows are
shown in the top panel of Figure \ref{fig:gradcont}, as blue and red
vectors, respectively.  We computed ${\cal M}[u_c], {\cal M}[u_g],
{\cal M}[u_c |d\Phi|/ \Phi_u],$ and ${\cal M}[u_g |d\Phi|/ \Phi_u].$
We also summed these quantities for each flow field to create
extensive measures of flows related to shearing and convergence.

We also computed PIL-weighted contour and gradient flows, ${\cal
M}[W_{PIL} u_c \,\tilde{B}_R ]$, and ${\cal M}[W_{PIL} u_c
\,\tilde{B}_R ]$, respectively.  The PIL-weighting function used here,
$W_{PIL}$, was meant to identify only regions of magnetograms near
PILs.  It was computed in the same way as the SPIL weighting function
$W_{SPIL}$ (which was used to determine $R$; see above), but with
different parameters.  The polarity bitmaps used for $W_{PIL}$ had a
threshold of only 40 G, while the threshold for the bitmaps used to
create $W_{SPIL}$ maps was 150 G; and the overlap-map of the dilated
polarity-mask bitmaps was convolved with a normalized Gaussian of
$\sim 7$ Mm FWHM for $W_{PIL}$, versus 15 Mm $W_{SPIL}$.  Weighting by
the signed magnetic flux, $d\Phi$, introduces a cross-PIL parity,
emphasizing regions with oppositely directed flows across the PIL, and
de-emphasizing regions where flows on opposite sides of the PIL are
similar.  Grayscale maps of $(W_{PIL} d\Phi)$-weighted gradient and
contour flows are shown in the left and right (resp.) bottom panels of
Figure \ref{fig:gradcont}.  In each of these bottom panels, contours
of $W_{PIL}$ for $\bar{B}_R$ from the top panel are overplotted in
red.  For some magnetograms, $W_{PIL}=0$ everywhere, and for some
others, $(W_{PIL} u) = 0$ everywhere, i.e., regions with $|\bar{B}_R|$
above the tracking threshold and non-zero PIL-map weighting do not
necessarily overlap.  We note that the product of a horizontal
velocity estimate with the radial magnetic field corresponds to a
horizontal electric field perpendicular to both; hence, the left and
right grayscale maps effectively show the components of the horizontal
electric field along contours and gradients (resp.) of $\bar{B}_R$.

\subsection{Expected Flow Properties}
\label{subsec:flow_expectations}

As an aside, we sought to confirm some expected correlations within
the dataset.  Of note, we expected diverging flows when flux was
emerging (see, e.g., \citealt{Abbett2000}), and converging flows when
flux was cancelling.  Assuming that flux emergence (cancellation)
leads to a positive (negative) $\dot \Phi$, we computed the linear and
rank-order correlations between FLCT's disk-passage averaged signed,
flux-weighted divergence and disk-passage averaged $\dot \Phi/\Phi$,
and found correlation coefficients of 0.61 and 0.67, respectively.
Using DAVE's disk-passage averaged signed, flux-weighted divergence in
the correlations gave values of 0.73 for both linear and rank-order
correlations.  We also computed the linear and rank-order correlations
between the center-of-flux divergence, $D_{COF}$, and $\dot
\Phi/\Phi$, and found correlation coefficients of 0.45 and 0.48,
respectively.

We also looked for evidence that shearing along PILs is associated
with flux emergence, as predicted by \citet{Manchester2007}.  First,
we correlated emerging flux with flux-weighted and unweighted measures
of curl, but coefficients were in the rage 0.2-0.4 --- smaller than
the correlations of emergence with divergence.  We also computed
rank-order and linear correlations between $S_{COF}$ and $\dot
\Phi/\Phi$ of 0.31 and 0.29, respectively; these are weaker than the
association of $D_{COF}$ with $\dot \Phi/\Phi$.  In addition, the
rank-order and linear correlations between disk-passage averaged, PIL-
and flux-weighted contour flows (see the discussion below equation
[\ref{eqn:conthat}]) and disk-passage averaged $\dot \Phi/\Phi$ were
0.13 and 0.36, respectively, for FLCT, and 0.24 and 0.42 for DAVE,
respectively.  Our measures of shearing motions were generally 
more weakly correlated with flux emergence than our measures of
diverging motions were.

Tests with synthetic magnetograms conducted by \citet{Welsch2008a}
with FLCT verified the supposition by \citet{Demoulin2003} that
estimated flow magnitudes $|u_c|$ along contours of $\bar{B}_R$ would
be underestimated in comparison to flow magnitudes $|u_g|$ along
gradients of $\bar{B}_R$.  We did not find, however, evidence of a
strong bias in flow speeds along and perpendicular to contours in our
data set: averaging over the $\sim 16.5$ million pixels for which both
FLCT and DAVE estimated flows, the mean and median contour speeds
$|u_c|$ were 2\% - 5\% higher than gradient speeds $|u_g|$ for both
FLCT and DAVE.  Differences between real and synthetic magnetograms
might explain the discrepancy between these results: real data are
more likely to possess fine-scale structure and asymmetries than
synthetic data, and these structural details make it possible for
tracking methods to determine the overall flow, even along contours.
It might still be the case, however, that our estimates are in fact
biased with respect to contours/ gradients in $\bar{B}_R$, and that
true contour speeds $|u_c|$ exceed true gradient speeds $|u_g|$ by an
even greater amount.

\begin{figure}
\plotone{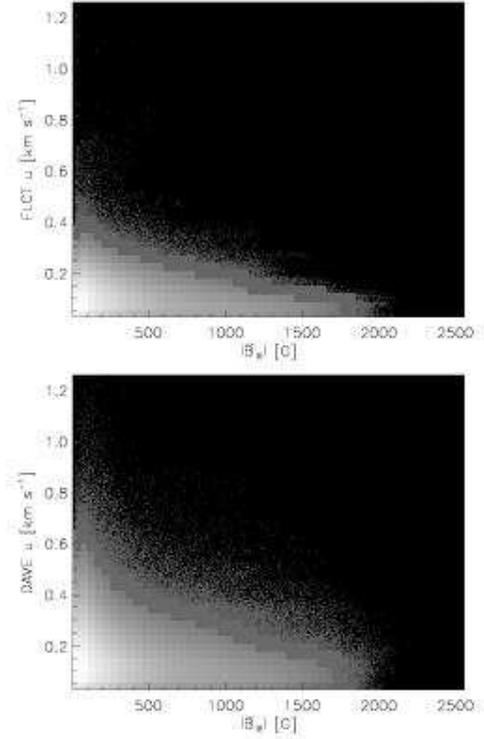}
\caption{%
Combined log-scaled shading histogram and scatter plot of speed
estimates from FLCT (top) and DAVE (bottom) as a function of
pixel-averaged field strength $\bar{B}_R$ (see text).  The grayscale
for FLCT ranges from 2.4 - 6.2, and for DAVE from 2.4 - 6.0.  Higher
speeds tend to be seen in weaker fields.
\label{fig:speedvb}}
\end{figure}

To understand the interplay between field strengths and estimated
speeds, we plotted the distributions of speeds $u$ with pixel-averaged
field strengths $\bar{B}_R$ for both FLCT and DAVE in Figure
\ref{fig:speedvb}.  The Figure includes speeds from all pixels for
which both FLCT and DAVE estimates were available.  Because these
several million speed estimates were too clustered for a regular
scatter plot, we binned speeds and field strengths, and used
log-scaled shading instead of points in high-density bins, and
overplotted individual points in low density bins.  The log-scale
shading for the FLCT plot ranges over 2.4 - 6.2, and for the DAVE plot
it ranges over 2.4 - 6.0.  Clearly, higher speeds tend to occur in
weaker fields, though a range of speeds are possible at any field
strength.  Statistically, however, correlations between speeds and
field strengths are weak, with rank-order coefficents for FLCT and
DAVE of 0.07 and -0.02, respectively.

\section{Associating Field \& Flow Properties with Flares}
\label{sec:associations}

\subsection{Disk-Passage Flare Associations}
\label{subsec:avg}

As a first step toward understanding the relationship between flows
and flares (if any), we investigated the correlations between each
AR's average GOES flux from flares of class C or greater, gross
magnetogram properties (e.g., $\Phi, R$, and their time derivatives),
and flow properties, averaged over the tracked component of the AR's
disk passage.  This coarse averaging in both space and time eliminates
many detailed aspects of magnetic field structure and evolution that
might be relevant to flare occurrence.  Nonetheless, any correlations
that remain despite this gross averaging ought to be robust.  The
values for flare flux used in these correlations are those listed in
the right-most column of Table \ref{tab:ar_data}.

\begin{figure}
\plotone{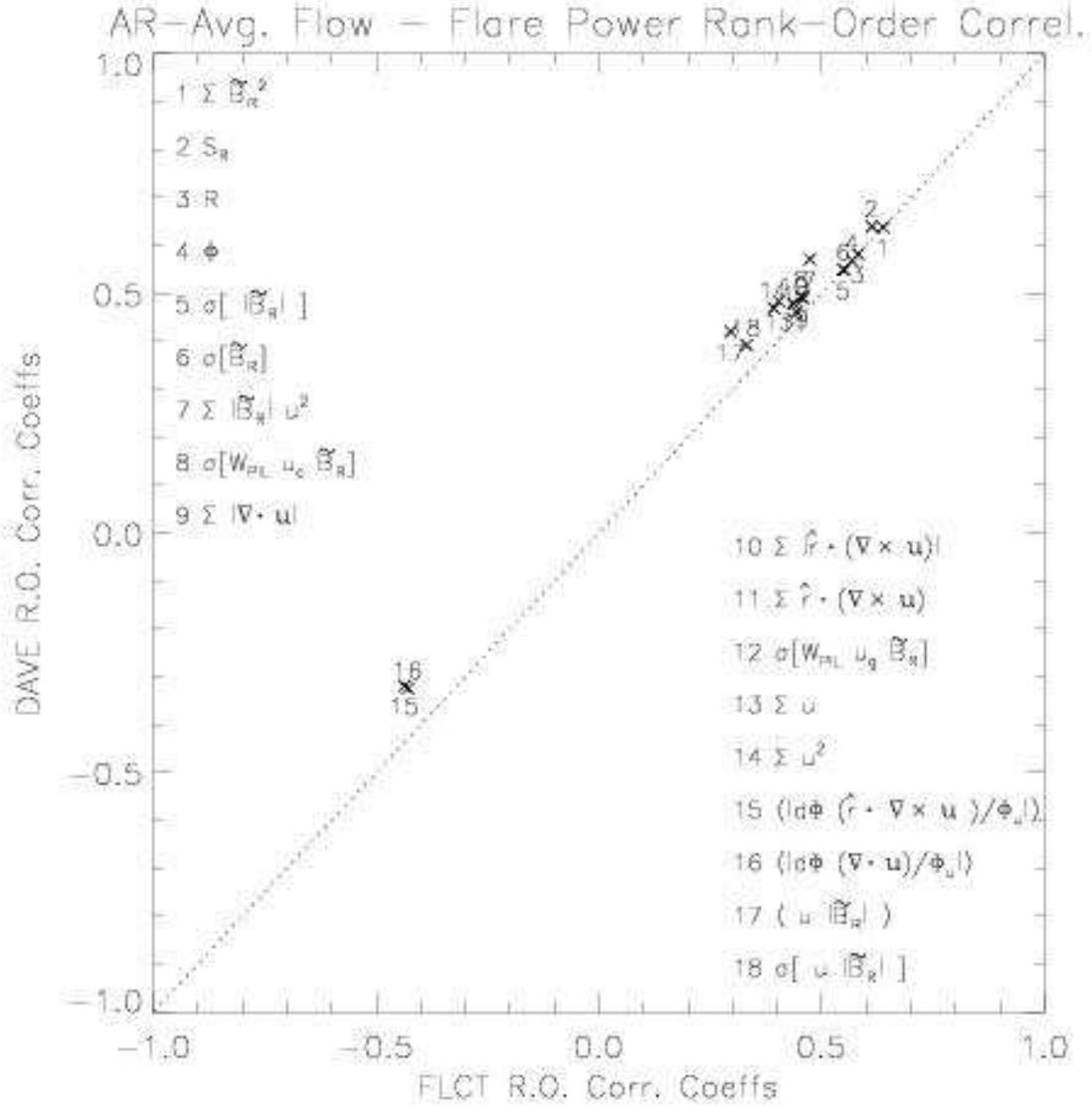}
\caption{%
FLCT and DAVE rank-order correlation coefficients between
whole-disk-passage averaged flow properties and flare power, averaged
over the 46 active regions in our study.  To the extent that
correlation coefficients lie on the slope $=1$ dashed line ({\em not}
a fit), the averaged flow properties from the FLCT and DAVE methods
agree.  The quantities correlated are numbered by distance from (0,0),
farthest first; hence, smaller numerical labels imply stronger
correlation.  Many of the strongest correlations do not require
estimating flows.  Even ranks are plotted above their symbols, and odd
ranks below.
\label{fig:flct_dave_ro}}
\end{figure}
\begin{figure}
\plotone{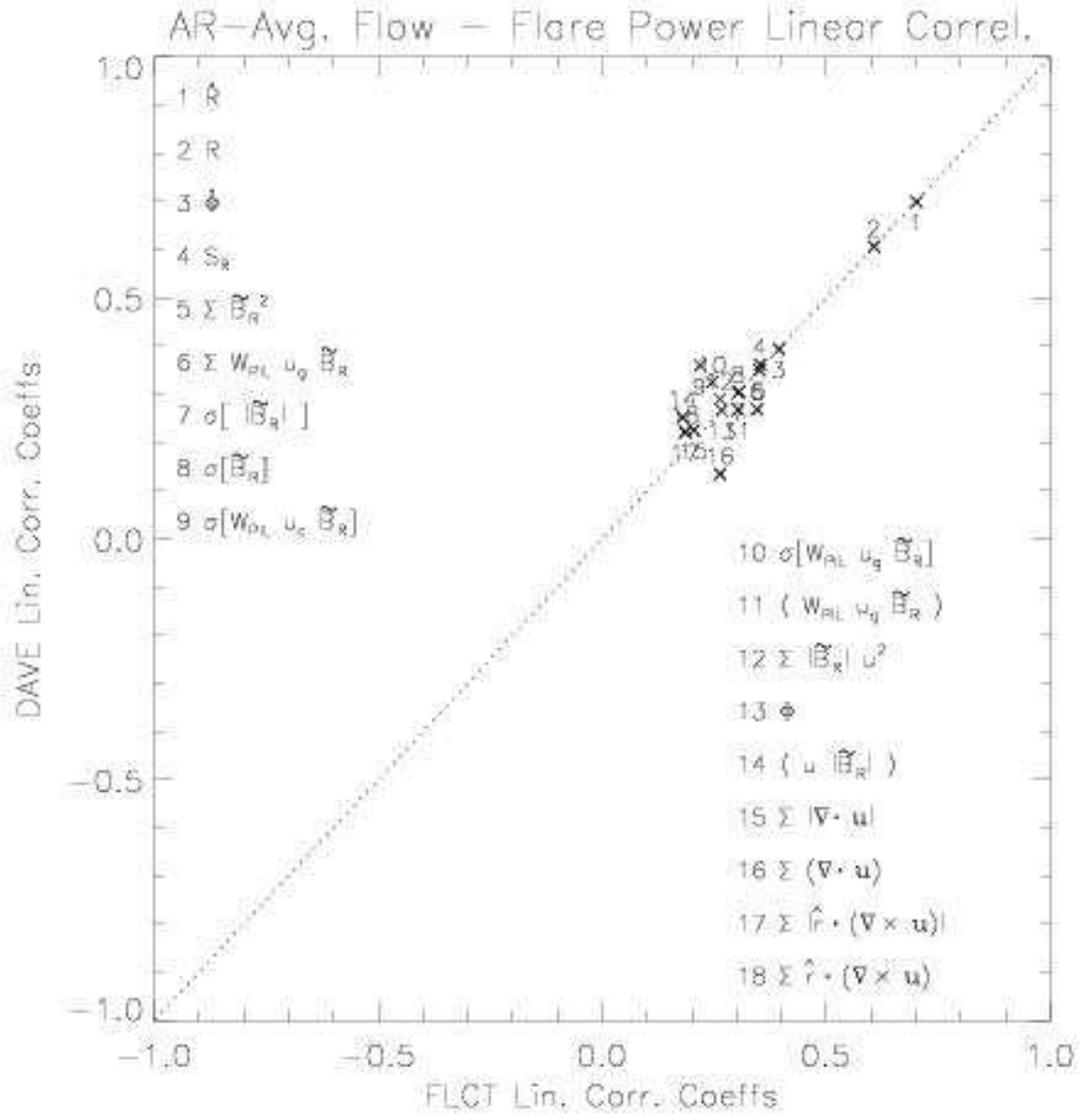}
\caption{%
FLCT and DAVE linear correlation coefficients between
whole-disk-passage averaged flow properties and flare power, averaged
over the 46 active regions in our study.  To the extent that
correlation coefficients lie on the slope $=1$ dashed line ({\em not}
a fit), the averaged flow properties from FLCT and DAVE methods agree.
The correlations are numbered by distance from (0,0), farthest first;
hence, smaller numerical labels imply stronger correlation.  As in
rank-order correlations in Figure \ref{fig:flct_dave_lin}, many of the
strongest correlations do not require estimating flows. Even ranks are
plotted above their symbols, and odd ranks below.
\label{fig:flct_dave_lin}}
\end{figure}

In Figures \ref{fig:flct_dave_ro} and \ref{fig:flct_dave_lin}, we
compare rank-order and linear (respectively) correlation coefficients
between whole-disk-passage averaged magnetic field and flow
properties, from both FLCT and DAVE, with whole-disk-passage averaged
flare power from equation (\ref{eqn:avg_power}), when averaged over
the 46 active regions in our study.  Quantities are labeled in order
of decreasing distance from (0,0), the point on these plots which
corresponds to complete lack of correlation with average flare power;
hence, smaller numerical labels imply stronger correlation.  Only
zeroth, first, and second moments of magnetic field and flow
properties are included in this ranking, and only the most strongly
correlated properties are shown.  (In plots that included higher
moments, no magnetic field properties were highly-ranked enough to
appear.  Some higher moments of {\em flow} properties, however, did
appear; but further analyses, described below, suggest that these
correlations were not robust.)  Odd rank numbers appear below their
corresponding plot symbols, while even rank numbers appear above their
symbols.  To the extent that flare correlations with a given flow
property from FLCT and DAVE flow agree more strongly, that flow
property's plotted symbol is nearer to the dotted diagonal line in
each plot (which is shown to guide the eye; it is not a fit).
Quantities that do not require a flow estimate (e.g., $R$) lie on the
diagonal line.  We believe this graphical presentation of our results
displays information more clearly than listing our results in tables
would.  Differences between linear and non-parametric, rank-order
correlation coefficients could arise from lack of linearity between
the correlated quantities, or non-normal variable distributions (e.g.,
numerous outliers), or our relatively low sample number $N_{AR} = 46$,
or some combination of these.  It is notable, for instance, that the
most highly linearly correlated quantity, $\dot R$, is not among the
most non-parametrically correlated quantities; but its ``excess kurtosis''
(that relative to the normal distribution) is 15, implying it deviates
very strongly from the normal distribution.

Some of the quantities most strongly associated with our measure of
disk-passage averaged flare flux, by either method of computing
correlations, do not require estimating flows: $\Phi, R, \sum
\tilde{B}_R^2,$ and quantities involving time derivatives of the first
two.  Some of these magnetic field - flare correlations accord with
previous work relating flare occurrence rates with total flux
\citep{Leka2007,Barnes2008} and flux near PILs
(\citealt{Schrijver2007}; and also for CMEs, as reported by Falconer
et al. 2003, 2006) \nocite{Falconer2003, Falconer2006}.  Table
\ref{tab:flux_corrs} shows that disk-passage averages of these
quantities are correlated with each other.  \citet{Schrijver2007} has
argued that large values of $R$ are more predictive of the occurrence
of large flares than $\Phi$ alone, but our correlation analysis does
not unequivocally support this assertion.  We note that the {\em
signed}, intensive rate of change of flux, $\dot \Phi/\Phi$, is more
strongly correlated with flare power than the unsigned intensive rate
of change, $|\dot \Phi|/\Phi$, implying that flux emergence is more
strongly correlated with flare energy release than flux cancellation
(which makes $\dot \Phi$ non-zero, but negative).  The extensive rate
of change of flux $\dot \Phi$ is more strongly correlated with average
flare flux than the intensive rate of change of flux, $\dot \Phi /
\Phi$, which suggests that more flare energy is released when more
photospheric flux emerges than when less flux does, even if the new
flux is larger relative to the pre-emergence flux.

Several extensive flow properties are significantly correlated with
our measure of disk-passage averaged flare flux.  Most notably, our
proxy Poynting flux, $S_R$, is among the quantities most strongly
correlated with average flare flux.

Some intensive flow properties show relatively strong negative
correlation with flare flux: the averages of field-weighted unsigned
vorticity, $|d\Phi (\hatr \cdot \nabla \times \uvec)|/\Phi_u$,
unsigned divergence, $|d\Phi (\nabla \cdot \uvec)|/\Phi_u$.
%
%
The magnetic flux - flare flux correlation is strong enough, however,
that normalization by $\Phi_u$ introduces a spurious, inverse
correlation between these field-weighted quantities and average flare
flux.  We found that the {\em unnormalized}, field-weighted versions
of these average flow properties (not shown) are {\em positively}
correlated with flare flux (but not sufficiently so to appear on these
plots of the most strongly correlated properties).

\begin{figure}
\plotone{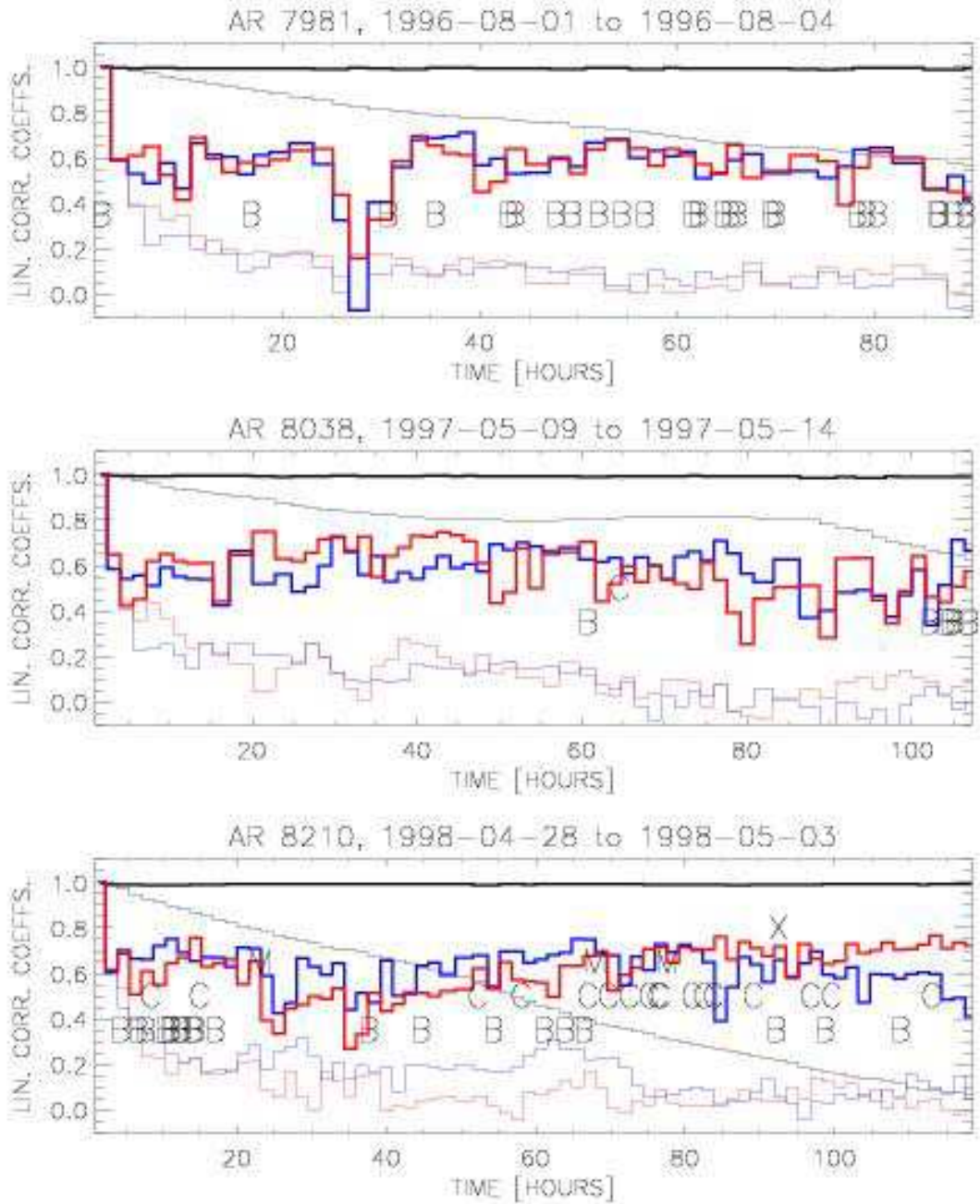}
\caption{%
For ARs 7981, 8038, and 8210, frame-to-frame correlations in
$\bar{B}_R$ with time are shown in black; the thick lines represent
linear correlation coefficients between the current and previous
frames, while the thin lines show linear correlation coefficients
between the current and initial frames.  Analogous correlations are
also shown for FLCT's estimated $u_x$ \& $u_y$ in blue \& red (resp.),
with thick \& thin lines denoting current-previous \& current-initial
correlations (resp.).  In addition, flare start times are shown by
letters representing GOES flare class; the vertical placements of
letters do not scale with actual flare fluxes.
\label{fig:tcorr_flct}}
\end{figure}

\begin{figure}
\plotone{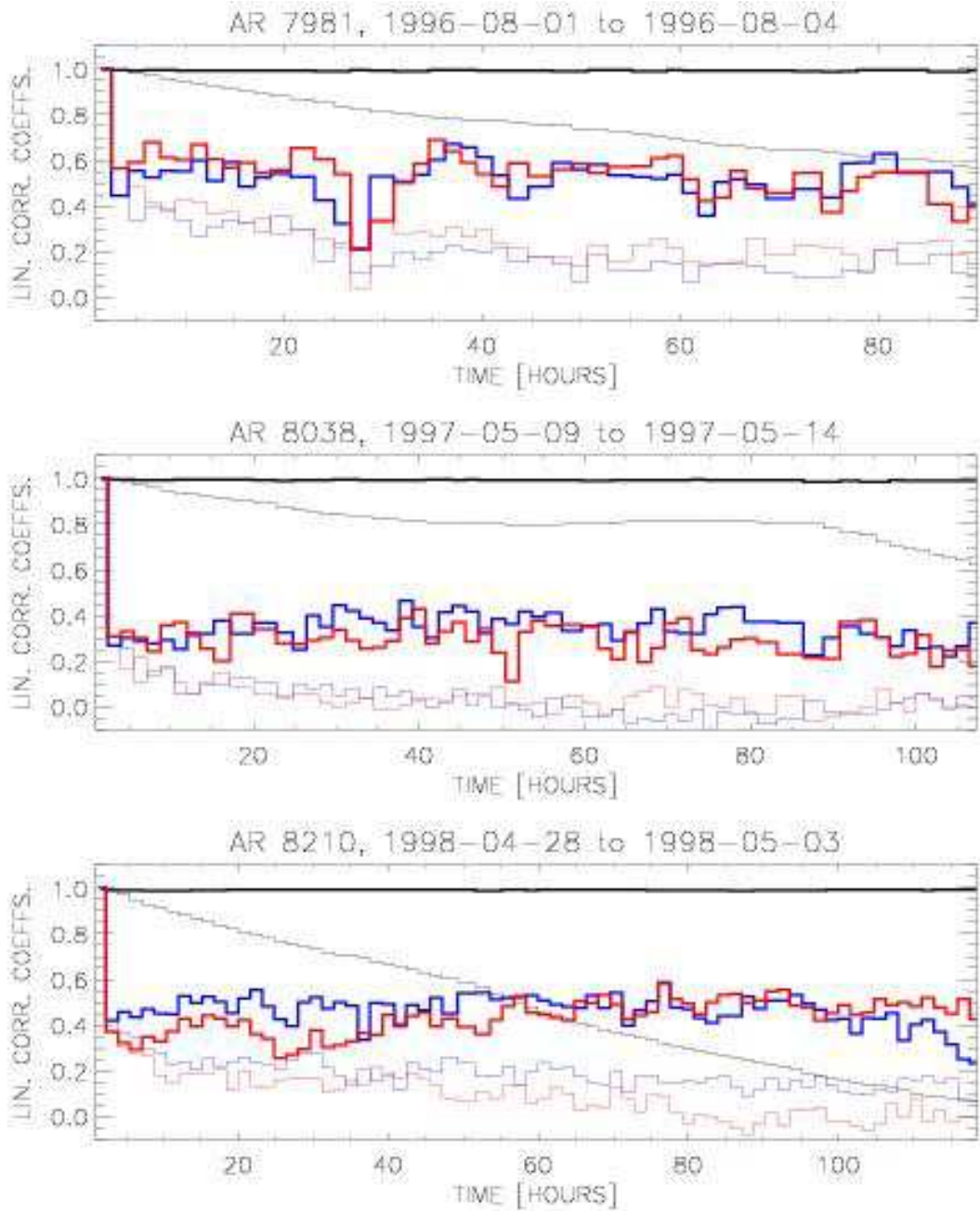}
\caption{%
As with Figure \ref{fig:tcorr_flct}, but with correlations using
DAVE's $u_x$ \& $u_y$ in blue \& red, resp.
\label{fig:tcorr_dave}}
\end{figure}

Apart from these flux-normalized flow properties, most other intensive
flow properties, averaged over the few days when ARs are near disk
center, are too weakly correlated with average GOES flux to appear in
these plots.  We expect, however, that AR magnetic fields are more
coherent on the relatively long timescale of a few days compared to
flows, which have lifetimes on the order of a few hours to a day
(e.g., \citealt{Leighton1962}).  Figures \ref{fig:tcorr_flct} and
\ref{fig:tcorr_dave} compare, for FLCT and DAVE, respectively, 
linear correlation coefficients of $\tilde{B}_R, u_x,$ and $u_y$ in
time --- both frame-to-frame ({\bf thick} black, blue, and red lines,
resp.) and with respect to the initial frame (thin black, blue, and
red lines, resp.)  --- for ARs 7981, 8038, and 8210.  Clearly,
frame-to-frame variations are much larger for estimated flows than for
$\tilde{B}_R$, meaning flow properties tend to average incoherently
--- toward zero --- in the disk-passage averaging we have undertaken
here.  This analysis suggests flows have a characteristic
decorrelation timescale --- taken as the time required for the
autocorrelation to fall to $1/e$ --- of $\sim 6$ hr. 

In Figure \ref{fig:tcorr_flct}, we also overplotted flare start times,
shown by letters representing GOES flare class. (Note that the vertical
placements of letters do not scale with actual flare fluxes.)  There
is no obvious correspondence between flare start times and field or
flow autocorrelations.  The differences between these three ARs'
current-frame-to-initial-frame magnetic field correlations are
intriguing, but investigating these variations is beyond the scope of
this study.

\subsection{Flare Associations on Shorter Timescales} 
\label{subsec:all}

\subsubsection{Correlation Analysis}
\label{subsubsec:corr}

The flow timescale implied by the foregoing correlation analysis ---
$e$-folding times of $\sim 6$ hr --- has implications for analyzing
flow-flare relationships.  Accepting the arguments of
\cite{McClymont1989}, we presume that flows do not directly power
flares, but still allow that some types of flows might tend to {\em
trigger} flares.  It is plausible that ``trigger'' flows have
the same characteristic timescale, $\sim 6$ hr, as the general flows
in our sample.  Alternatively, if some types of photospheric flows
correspond to the injection of energy (perhaps via flux emergence)
into the corona over several hours preceding flares (thereby being the
{\em ultimate} source of flare energy, as opposed to the {\em
proximate} energy source if flows were to drive flares), then there
ought to be a flow-flare relationship on the coronal energy storage
timescale.  Work by \citet{Schrijver2005} and \citet{Longcope2005}
suggests the timescale for coronal relaxation via flaring and
reconnection is on the order of 24 hr.  Accordingly, we now address
the relationship between flow properties and the occurrence of flares
in these two shorter time windows, 6 and 24 hours.

As a first step, we correlated magnetogram and flow properties with
the average GOES flux from C-class and larger flares, computed via
equation (\ref{eqn:avg_power}), but with the disk-passage time
interval $\tau_{45}$ replaced with the flare window length, $\tau_6$
for 6 hours, or $\tau_{24}$ for 24 hours.  For each possible window
duration, two intervals were analyzed: (1) the ``current'' window,
centered on the average time of the two magnetograms used to estimate
flows; and (2) the ``next'' window of the same duration following the
current window.  For ease of reference, we label these four windows
6C, 24C, 6N and 24N, with C and N denoting the current and next
windows.  To be explicit, our 6C window includes flares in the 6-hour
interval $\pm 3$ hours from each velocity estimate, and 6N includes
flares in the 6-hour interval 3-9 hours after that velocity estimate.
The 24C window includes flares within $\pm 12$ hours of each velocity
estimate, and 24N includes flares over the 24-hour interval between
12-36 after that velocity estimate.  In space weather parlance,
forecasts corresponding to the 6N and 24N windows would have {\em
latencies} of 3 and 12 hours, respectively.  Note that our time
windows do not correspond to the zero-latency forecast windows used by
some other authors, e.g., \citet{Barnes2007}.

\begin{figure}
\plotone{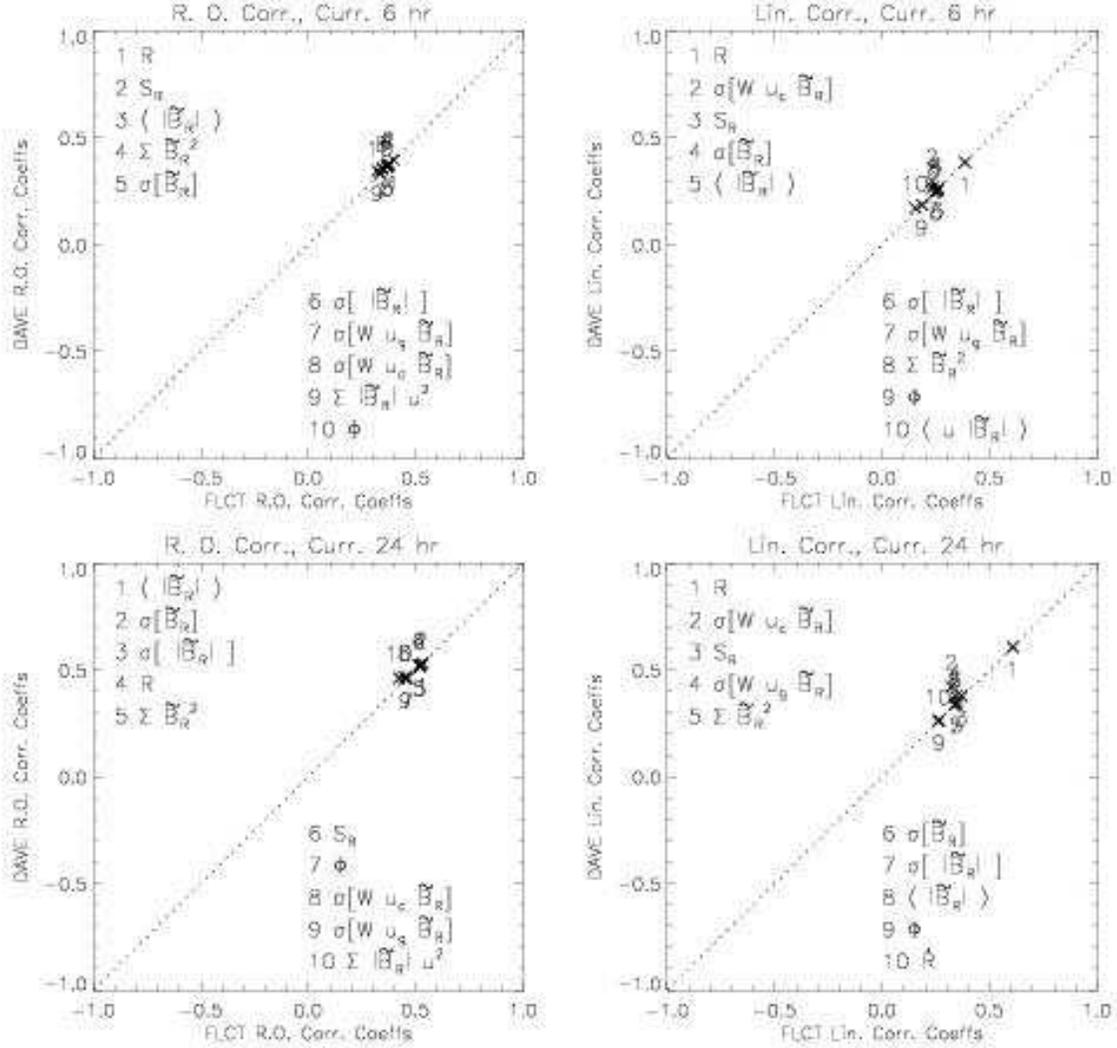}
\caption{%
%
Top row: Scatter plots of rank-order (R.O.) and linear correlation
coefficients (left and right panels, resp.)  between average flare
power and magnetogram or flow properties from FLCT (horizontal axis)
and DAVE (vertical axis), within the 6-hour time window, 6C, centered
on the average time of the magnetograms used to estimate flows.
Quantities are labeled in order of decreasing distance from (0,0),
the point on the plot which corresponds to complete lack of
correlation with average flare power.  Low-ranked properties are not
shown.  Odd rank numbers appear below their corresponding plot
symbols, while even rank numbers appear above their symbols. To the
extent that flare correlations with flow properties from FLCT and DAVE
flow agree, they lie along the dotted line (not a fit).  Quantities
that do not require a flow estimate (e.g., total flux, $\Phi$) lie on
the diagonal line.  Bottom row: same as for top, but for the 24-hour
time window 24C.  Some of the quantities most strongly correlated with
average flare power --- e.g., $R$ --- do not require estimating flows.
Our proxy for the Poynting flux, $S_R$, exhibits comparable
correlations.
%
%
%
\label{fig:all_curr}}
\end{figure}

\begin{figure}
\plotone{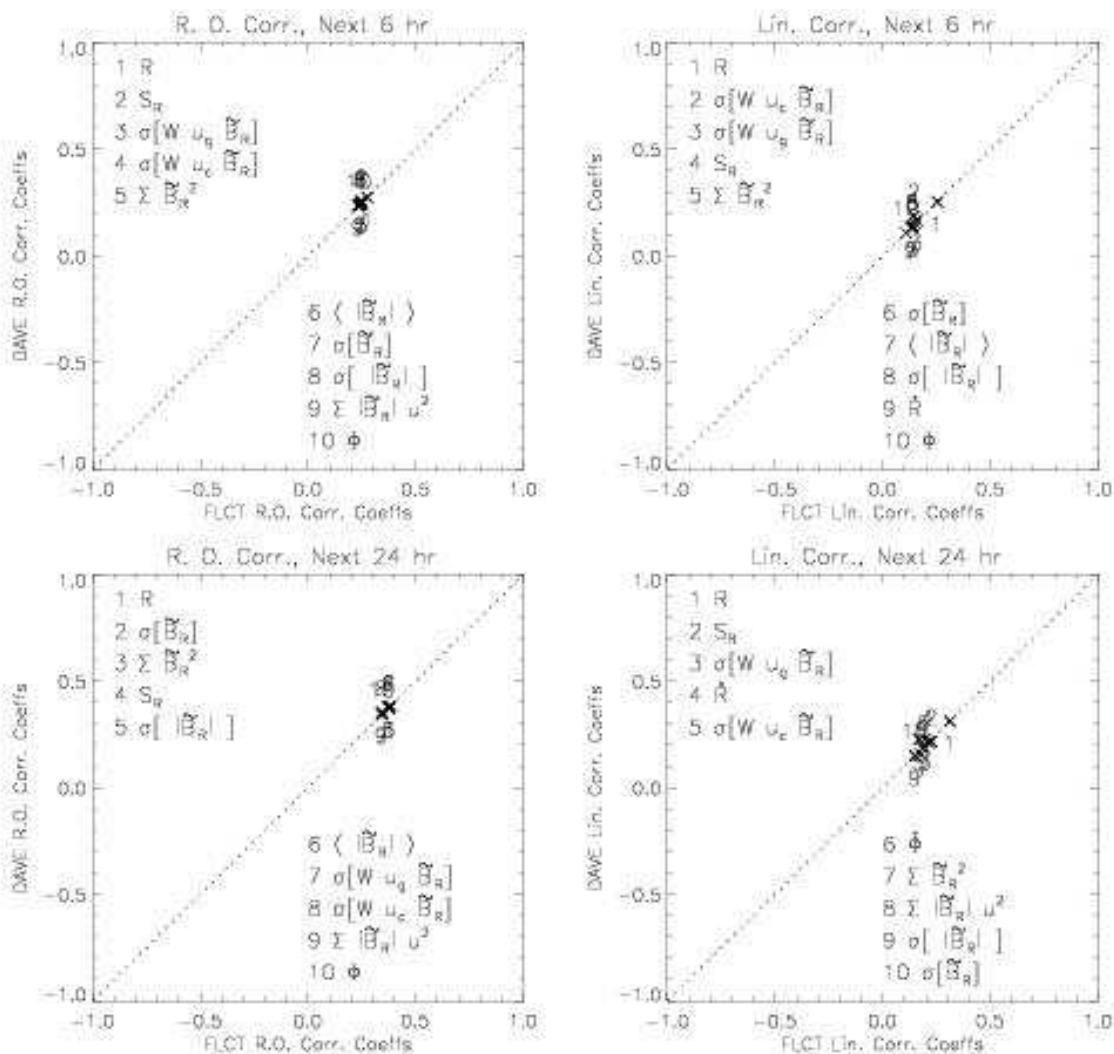}
\caption{%
As with Figure \ref{fig:all_curr}, these plots show correlation
coefficients between properties of magnetograms or flows from FLCT
(horizontal axis) and DAVE (vertical axis), and average flare power
from the 6N (top row) and 24N (bottom row) windows. In each panel,
quantities are labeled in order of their distance from (0,0).
Low-ranked properties are not shown.  Odd rank numbers appear below
their corresponding plot symbols, while even rank numbers appear above
their symbols.  As with Figure \ref{fig:all_curr}, many quantities
strongly correlated with average flare flux do not require estimating
flows, but our proxy for the Poynting flux, $S_R$, exhibits comparable
correlations.
\label{fig:all_next}}
\end{figure}

The scatter plots in Figures \ref{fig:all_curr} and \ref{fig:all_next}
show rank-order (R.O.) and linear correlation coefficients, for both C
and N time windows, between average flare power and magnetogram or
flow properties from FLCT (horizontal axis) and DAVE (vertical axis).
The format of plots in these Figures is similar to that of Figures
\ref{fig:flct_dave_ro} and \ref{fig:flct_dave_lin}.  Correlated
quantities are labeled in order of decreasing distance from (0,0),
the point on the plots which corresponds to complete lack of
correlation with average flare power.  Odd rank numbers appear below
their corresponding plot symbols, while even rank numbers appear above
their symbols.  In these plots, {\em all} moments of magnetogram and
flow properties were ranked, and only the ten most highly ranked
properties were shown.  To the extent that flare correlations with
flow properties from FLCT and DAVE agree, they lie along the diagonal
dotted line (not a fit).  Quantities that do not require a flow
estimate (e.g., $R$) lie on the diagonal line.  No higher moments of
magnetic field or flow properties beyond the zeroth, first, and second
were correlated strongly enough with flare flux in these windows to
appear in these plots.

These plots show relationships similar to the disk-passage averaged
results.  As with the disk-passage averaged correlations, many of the
quantities that are strongly positively correlated with average flare
power do not require estimating flows --- e.g., $\Phi, R, \sum
\tilde{B}_R^2$, the average unsigned field $\lan
|\tilde{B}_R| \ran$, as well as the standard deviations of the signed
and unsigned fields, $\sigma[\tilde{B}_R]$ and
$\sigma[|\tilde{B}_R|]$, respectively.  Again, our proxy for the
Poynting flux $S_R$ is among the quantities most strongly correlated
with average flare flux.

The coefficient of determination is defined as $r^2$, where $r$ is the
linear correlation coefficient.  It quantifies how much of the
variation in the dependent variable (avg. flare flux, here) is
explained by the independent variable: when $r^2 = 1,$ all of the
observed variation can be explained in terms of the correlation, and a
smaller value of $r^2$ implies less explanatory power.  The largest
values of $r^2$ for the correlations with flare flux that we performed
are $\sim 0.25$, implying most variation in flare flux cannot be
explained by any single variable that we considered.   

The clustering of correlation coefficients from magnetic field and
flow properties in Figures \ref{fig:all_curr} - \ref{fig:all_next}
hampers clear identification of which magnetic field and flow
properties are most relevant to flare activity.  Hence, relationships
between these properties and flare activity should be analyzed
further.

\subsubsection{Discriminant Analysis}
\label{subsubsec:da}

We also used the discriminant analysis (DA) code developed by Barnes,
Leka, and collaborators (see, e.g.,
\citealt{Barnes2007,Leka2007,Leka2003b}) to analyze our dataset.
Briefly, given samples classified into distinct groups (in our case,
time windows that have been classified into binary populations, either
flare-associated or flare-quiet) and a set of observed quantities
corresponding to each sample (in our case, magnetic field and flow
properties), DA can be used to generate a discriminant function --- a
linear combination of the observed quantities --- that, in a
particular sense, best reproduces the classification of each sample
into the distinct groups.  This discriminant function can then be used
as a way to predict the class to which a future sample will belong.
While its approach of determining the linear combination of observed
quantities that can explain variation in the samples is similar to
that of principal component analysis, DA focuses on determining which
observed quantities most strongly distinguish between the classes.

In principle, with the observed quantities standardized by subtracting
their means and dividing by their variances (using, e.g., IDL's
\textsc{standardize.pro}), ratios of coefficients in the discriminant function
quantify the relative discriminatory power of the observed quantities.
Thus, DA might be used to identify the strongest discriminators, even
if several of the observed quantities are correlated with each other.
Values of these coefficients can, however, be unstable (e.g.,
\citealt{Huberty1992}), implying such ratios cannot be used to rank
the importance of observed quantities with coefficients of similar
magnitude.  Nonetheless, DA should still be able to distinguish
between variables whose coefficients differ significantly.  A variable
with a much smaller discriminant coefficient either inherently lacks
power to accurately classify a sample, or contributes little
additional discriminatory power beyond that from another variable in
the discriminant function with which it is highly correlated.  In
accordance with these considerations, we focus less on ranking the
utility of variables, versus identifying the few with strong
discriminatory power in each time window.
 
Among other things, DA assumes the distribution of each observed
quantity is normal; but \citet{Leka2007} found the results of DA
applied to vector magnetogram properties as predictors of solar flare
occurrence to be relatively insensitive to the distributions of input
properties.  In addition, DA assumes that the observed quantities are
independently measured.  In our study, this is not strictly true for
magnetogram properties like $\Phi, R,$ and $\sum \tilde{B}_R^2$, which
were measured from the estimated radial field $\bar{B}_R$ averaged
from successive magnetograms; these quantities calculated from
successive magnetogram pairs are not formally independent.  The high
frame-to-frame correlations for the AR magnetic fields shown in Figure
\ref{fig:tcorr_flct}, however, imply that even if formally independent
magnetogram measurements had been used, the resulting magnetic field
properties would still be highly correlated.

We used DA to investigate the relationship between estimated
magnetic field / flow properties and the occurrence of C-class (or
stronger) flares, in each of the four time windows (6C, 6N, 24C, 24N)
associated with each estimation of magnetic field and flow properties.
With the nominal 96-minute cadence of estimates of magnetogram and
flow properties in our study, several estimates might be associated
with a single flare.  Nonetheless, DA should still be able to quantify
the relative power of observed magnetic field and flow properties to
discriminate between flare-quiet and flare-associated windows.

While DA can identify the subset of variables with the most
discriminatory power, the predictive power of that combined set of
variables (with relative weights given by the discriminant function)
must be evaluated separately.  Climatological skill scores (see, e.g.,
\citealt{Wheatland2005}), which quantify the improvement in forecast
accuracy over the constant forecast derived from the average event
frequency, can be used for this purpose.
After \citet{Wheatland2005}, we denote averaging over all samples by
$\lan ... \ran$, the forecast probability by $f$, and the actual
occurrence by $x$ (with a value of 1 or 0 for each sample, for flaring
or not, respectively), with a skill score $SS$ given by
\be SS = 1 - \lan (f - x)^2 \ran / \lan (\lan x \ran - x)^2 \ran
~. \label{eqn:ss} \ee 
Perfect forecasting results in $SS = 1$, and scores below zero reflect
worse performance than a uniform ``climatological'' prediction of
flare likelihood equal to the observed frequency $\lan x \ran$.  While
several other metrics of forecast models' predictive power exist,
validation of a forecast model is not the object of this paper, so we
opted not to consider additional metrics.

Which properties associated with each pair of magnetograms had the
greatest discriminatory power for each time window?  We note that, in
addition to the magnetic field and flow properties considered in the
correlation analysis above, we also considered the average flare flux
during the current time windows, 6C and 24C, as a property that could
be included in discriminant functions for windows 6N and 24N,
respectively.  Because $S_R$ was strongly correlated with average
flare flux for many time windows in the analysis above, we started by
using DA to construct separate, two-variable discriminant functions
with $S_R$ and each of the other possible magnetic field, flow, and
flare (for windows 6N and 24N) variables in turn, for each of the four
time windows.  Separate versions of these discriminant functions were
calculated using FLCT and DAVE flow properties.  In this way, each
property was separately tested ``head to head'' against $S_R$.  For
some windows and/ or flow estimates, it was found that $R$ (and only
$R$) had a larger discriminant coefficient than $S_R$.  Accordingly,
for these windows, we repeated this head-to-head analysis,
constructing two-variable discriminant functions with $R$ and each of
the other variables in turn; and no other variable had a larger
coefficient in the two-variable discriminant functions for these
windows than $R$.  These computations also identified the
quantity that, for each window, had the most additional discriminatory
power: the quantity with a discriminant coefficient closest in
magnitude to the highest.  For all windows but 24C, $S_R$ and $R$ had
the largest coefficients.  For 24C, $\lan |\tilde{B}_R| \ran$ --- the
average absolute field strength, an instrinsic magnetic quantity ---
had the largest coefficient relative to $S_R$.

We then constructed three-variable discriminant functions, separately
for each additional magnetic field and flow property (also separately
for FLCT and DAVE flows) beyond the two properties already identified
for each time window.  At this stage, the hierarchy of discriminant
function coefficients differed between the two flow estimation methods
for all windows, though skill scores only differed by at most $0.02.$
As additional variables beyond these were considered, skill scores did
not substantially increase, and the hierarchies of discriminant
function coefficients constructed using variables from the two
different flow estimates methods continued to disagree.  Both of these
facts suggested that including additional variables would result in
diminishing marginal improvement in identifying magnetic field and
flow properties associated with flares.  Accordingly, we did not
pursue this analysis further.

Table \ref{tab:skills} summarizes the outcome of our discriminant
analysis.  Time windows are listed in the left-most column, and method
used to estimate flows are listed in the next column.  The variables
used to construct each discriminant function are shown in the next
column.  For multiple-variable discriminant functions, the
discriminant coefficients of each standardized variable are listed in
square brackets.  These results suggest that $S_R$ and $R$
consistently have strong discriminatory power between flaring and
non-flaring populations, regardless of tracking method used to
estimate $S_R$.  Further, given their commensurate discriminant
coefficients, our discriminant analysis implies that these two
parameters each possess some unique predictive capability.  The
(intensive) average unsigned field strength, $\lan |\tilde{B}_R|
\ran$, also had relatively large discriminatory power in many cases.

The next four columns of Table \ref{tab:skills} show predicted and
actual flare frequencies for each window considered.  (There are many
more flares in the statistics listed in these columns --- from around a
hundred to a few hundred, depending on the window --- than are listed
in \S \ref{sec:data}.  This discrepancy results from ``overlapping
measurements'': many average magnetic fields and flow estimates having
a common flare in their windows.)  In classifying events for each
window, we used the same data that were used to construct the
discrimant function for that window, which tends to bias the success
rate upward \citep{Hills1966}.  (Unbiased calculations require
constructing separate discriminant functions with each sample removed,
and using the resulting discriminant function to classify the removed
sample.  This is both time consuming, and, in tests we performed, did
not significantly alter event classifications.)  The final column
lists the climatological skill scores, $SS$, derived for the
corresponding discriminant function.  For comparison with the skill
scores reported here, \citet{Barnes2007} used a large set of
properties of AR magnetic fields derived from vector magnetograms, and
achieved an $SS$ value of 0.35 for predicting C-class or larger flares
in the 24 hours following each magnetogram.  Our differing definition
of the future 24 hour window (12-hour latency in our case, no latency
in theirs) precludes direct comparison of our results with theirs, but
their results nonetheless provide useful context.  

We also note that the vector magnetograms used by \citet{Barnes2007}
are probably superior to LOS magnetograms for computing radial
magnetic field strengths.  For instance, \citet{Knoll2008} computed
the skill scores for single-variable discriminant functions using
total unsigned flux, computed from cosine-corrected LOS magnetograms
at various distances from disk center; they found that the skill score
using magnetograms within 45$^\circ$ of disk center (as with our data
set) was a factor of three lower than the skill score using only
magnetograms from within 10$^\circ$ of disk center.  This demonstrates
that using off-center LOS data to estimate $B_r$, as we have done, can
negatively impact skill scores.

Skill scores were negative for the 6N window using our discriminant
functions, implying no skill beyond that based upon event frequencies
in the data set.  As flares are infrequent, forecasting for small time
windows tends to be inaccurate, as a short time interval is more
likely than not to be flare quiet \citep{Leka2007}; this might partly
explain why forecast skill for window 24N exceeded that of 6N.  For
window 24N, flare flux averaged over the current 24 hour window,
${\cal F}_{24,{\rm current}}$, exhibited some discriminatory power,
and evidently shares some predictive power with $R$, as the latter's
coefficient shrank when the former was included in the discriminant
function.  The tendency for regions that have already flared to flare
again soon after is referred to as persistence in the flare forecast
literature \citep{Zirin1991}.

In Table \ref{tab:popstats}, we list statistical properties of the two
variables for each time window with the highest discriminant function
coefficients from Table \ref{tab:skills}.  We first list the means of
the raw variables for the flaring and non-flaring populations from
each window, separately and combined; and then list the corresponding
standard deviations of each variable's combined flaring and
non-flaring populations. We then list the means of the standardized
versions of each variable for the flaring and non-flaring populations.
Results from FLCT were used for the only quantity that involves flows,
$S_R$.

\section{Discussion \& Conclusions}
\label{sec:conc}


Magnetic energy is the predominant form of energy in the low corona,
and free magnetic energy is thought to power flares and CMEs
\citep{Forbes2000}.  This free energy is thought to emerge into the
corona in active region magnetic fields \citep{McClymont1989}.  As the
photospheric magnetic field should be strongly coupled to the plasma
(e.g., \citealt{Parker1984a}), the transport of magnetic energy across
the photosphere should be governed by equation
(\ref{eqn:3comp_poynting}): either radial (vertical) flows transport
current-carrying magnetic flux across the photosphere (e.g.,
\citealt{Leka1996,Abbett2003}), or photospheric flows act on already
emerged fields (e.g., \citealt{Brown2003,Antiochos1999a}).  Accurate
estimates of photospheric flows, therefore, could be used to study
physical processes that lead to the buildup of coronal magnetic energy
that is released in flares and CMEs.  If processes that trigger
flares/ CMEs are related to photospheric evolution, then studying
photospheric flows can improve our understanding of these processes.
It is possible, however, that trigger processes are essentially
coronal, and only related to photospheric evolution in a statistical
sense.  In the latter case, knowledge about trigger processes that
could be gleaned from studies of photospheric evolution would be
limited.

In this study, we attempted to relate properties of magnetic fields
and flow fields estimated by tracking photospheric magnetograms to the
release of coronal magnetic energy in the form of flares.  We
distinguished between extensive properties, which scale with AR size,
and intensive properties, which do not depend directly upon AR size.
Using correlation analysis, we identified several properties that were
roughly equally associated with flare flux averaged over several time
windows.  To distinguish which among these properties was most
strongly related to the occurrence of flares, we employed linear
discriminant analysis.  Using both methods, we found one extensive
flow property, $S_R = \sum u \tilde{B}_R^2$, to be associated with
flare flux and flare occurrence at a level comparable to the most
strongly flare-associated magnetic field property, $R$, which
quantifies the flux near strong-field polarity inversion lines
(SPILs).  Previous work by Falconer et al. (2003, 2006)
\nocite{Falconer2003,Falconer2006} and \citet{Schrijver2007} had
already linked the amount of flux near SPILs to CMEs and flares.

The association of $S_R$ with flare activity was robust
regardless of whether FLCT or DAVE were used to estimate flows, and
for a variety of time windows corresponding to each flow estimate.  To
quantify the similarity between the flows when weighted by
$\tilde{B}_R^2$, we computed the linear and rank-order correlation
coefficients between $u_{\rm FLCT} \tilde{B}_R^2$ and $u_{\rm DAVE}
\tilde{B}_R^2$ for the $\sim 16.5$ million pixels where both estimated
flows; they were 0.89 and 0.91, respectively.  We note that $S_R$ was
more strongly associated with flares than $\sum \tilde{B}_R^2$,
implying the square of the magnetic field strength is not solely
responsible for association.  We also found $u$ and $|\bar{B}_R|$ to
be uncorrelated with each other, so the association with
flaring does not arise from a hidden correlation between $u$ and
$|\bar{B}_R|$.  Hence, we conclude that flow information is an
essential factor in the association between $S_R$ and flare activity
that we have identified.  We could not unambiguously associate any
intensive property of flows with flare activity, a potentially
significant fact we discuss in more detail below.

The extensive or intensive scaling of flare-associated magnetic field
and flow properties is relevant to flare and CME processes.  Triggers
of flares and CMEs might be small-scale features like current sheets;
for instance, globally significant reconnection begins at current
sheets in the breakout model of \citet{Antiochos1999a} and the tether
cutting model of \citet{Moore2001}.  In contrast, the non-potential
magnetic structures associated with flares and CMEs, such as filaments
and sigmoids, are large-scale features.  Free magnetic energy is
probably stored on large scales, because the difference between the
actual and potential magnetic fields is non-local.  For example, the
presence of a current sheet --- a compact non-potential structure ---
can alter the magnetic field far from the current sheet itself.
Flares occur over a range of energies (e.g., \citealt{Hudson1991}),
and larger ARs can produce larger flares \citep{Sammis2000,
Kucera1997}, suggesting that peak flare energy is an extensive
property of ARs.  The effectively exponential weighting of average
flare flux by flare class in equation (\ref{eqn:avg_power}) means that
the average flare fluxes we used here strongly emphasize the largest
flares.  Hence, our measure of flare activity essentially scales
extensively.  Consequently, it is perhaps unsurprising that the AR
properties we found to be most strongly associated with flare activity
are extensive.

\begin{figure}[htp]
\plotone{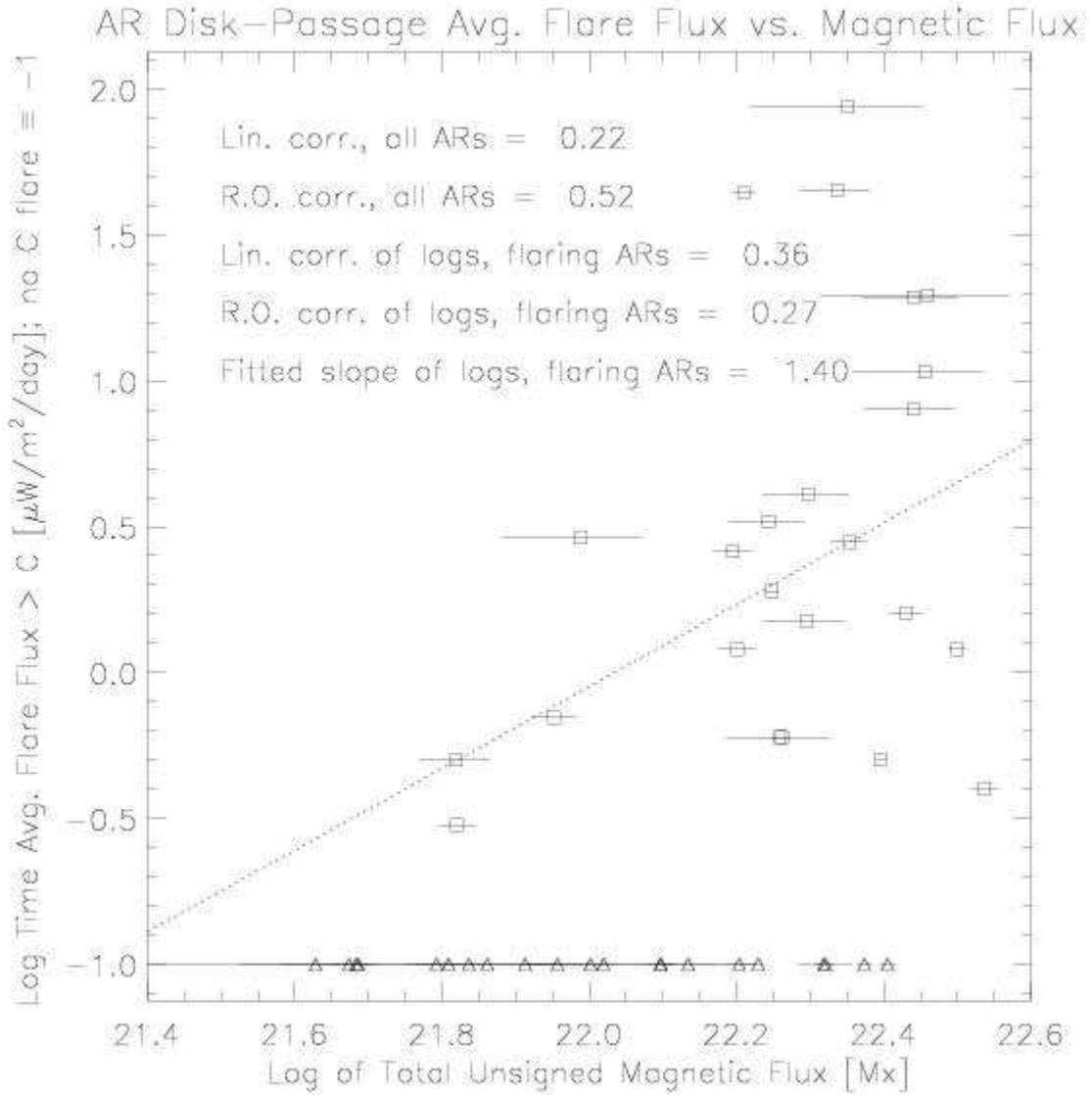}
\caption{%
Squares show $\log_{10}$ average flux from flares above GOES C1.0
level while each AR was within 45$^\circ$ of disk center vs. log
average total unsigned flux in these ARs over that time interval.
Horizontal bars show log of RMS variations flux during that interval;
errors from magnetograph noise for any single flux measurement are
about the size of the plotted symbols or smaller.  Triangles show log
of magnetic flux in the 22 ARs that did not flare above GOES C1.0
level.  We also list: the linear and rank-order (R.O.)  correlation
coefficients between the raw magnetic and flare fluxes for all ARs;
the linear and R.O. coefficients between the logs of magnetic and
flare fluxes for flaring ARs; and the slope of a least- absolute-
deviation fit (dotted line) of the log of flare flux as a function of
log of magnetic flux for ARs that flared above the C1.0 level.
\label{fig:flux_vs_flux}}
\end{figure}

Figure \ref{fig:flux_vs_flux} illustrates this extensive scaling.  The
squares plot the relationship between the log of the average flux from
flares at or above GOES C1.0 level over the time interval $\tau_{45}$
when each AR was within 45$^\circ$ of disk center, versus the log of
the average total unsigned magnetic flux in these ARs over that time.
Error bars show RMS variance in flux for each AR over $\tau_{45}$;
errors expected from 20 G per pixel noise in an individual MDI
magnetogram are about the size of the plotted squares or smaller.
Triangles show the flux in the 22 ARs that did not flare at or above
GOES C1.0 level.  (Because many A- and B-class flares are not reliably
ascribed to a source AR in the GOES catalog, emission from these
flares has been excluded.)  It is clear the squares show a trend,
which we crudely quantified by computing linear and rank-order (R.O.)
correlations between magnetic flux and average flux from flares above
the GOES C1.0 level, both for raw values using all ARs, and for
logarithms of these values using only ARs that flared above this
level.  We also computed a least- absolute- deviation fit of the log
of average flare flux as a function of the log of unsigned magnetic
flux, which yielded a slope of 1.40.  While our small sample size
hampers our ability rigorously characterize this relationship, we note
that the value of the slope differs from the value of $\sim 1.15$
found by \citet{Pevtsov2003} to relate total unsigned flux to thermal
soft X-ray (SXR) luminosity.  A discrepancy between the scalings of
thermal and non-thermal (flare) SXR luminosities with AR magnetic flux
might indicate that different physical processes generate each type of
emission.  Also, the magnetic flux - flare flux correlation might
account for some of the variability in AR flare productivity reported
in other studies relating flares to magnetic field properties and/ or
evolution --- e.g., \citet{Abramenko2005} and \citealt{LaBonte2007}.
An intensive measure of AR flare productivity might be formed by
normalizing average AR flare flux by AR magnetic flux $\Phi$
raised to some power $a$; for a larger sample, one could follow the
procedure adopted by \citet{Fisher1998} to determine an exponent that
removed any correlation between average flare flux and magnetic flux.
Linearity in the magnetic flux - flare flux relationship could be
consistent with uniform free energy input per unit flux (uniform
``specific free magnetic energy'').  Nonlinearity might reflect
magnetic interactions after emergence (as would occur in an avalanche
model of flares, \citealt{Lu1991}), or an aspect of the processes
(dynamo, emgergence) that produce active region magnetic fields.

Given $R$'s extensive scaling, it is interesting that
\citet{Schrijver2007} found that ARs with similar values of $R$ could
produce flares with a range of energies.  This suggests that the
amount of flux near SPILs does not directly imply the amount of energy
that will be released in any subsequent flares.  It is possible that
$R$ scales with the total amount of magnetic energy available to be
released, but that the size of the actual energy released depends upon
the details of the coronal magnetic field configuration.
\citet{Welsch2008b} found that $R$ tended to increase while total AR
flux was increasing, implying that concentrations of flux near SPILs
tend to arise by the emergence of flux, as opposed to the convergence
of already emerged flux.  Emergence has long been supposed to be a
trigger of flares (e.g., \citealt{Priest1974}).  So, despite its
extensive scaling, a large value of $R$ might indicate the presence of
a flare trigger, as well as the total amount of free magnetic energy
present in a configuration.

The flow property most strongly associated with average flare flux,
$S_R$, has units of an energy flux: energy per unit area per unit
time.  In Table \ref{tab:popstats}, raw values for this quantity are
$\sim 10^7$ km G$^2$ sec$^{-1}$; mulitplying by both $10^5$, to
convert km to cm, and the MDI disk-center pixel area, $da \simeq 2
\times 10^{16}$ cm$^2$, and dividing by $4 \pi$ (for consistency with
the Poynting flux in equation [\ref{eqn:true_poynting}]) gives a power
on the order of $10^{27}$ erg sec$^{-1}$.  Integrated over the $\sim
10^5$ seconds in a day or so, this represents the transport of $\sim
10^{32}$ ergs day$^{-1}$ of magnetic energy.  While we have referred
to $S_R$ as a proxy for the integrated Poynting flux of magnetic
energy across the photosphere, actual correlation between these
quantities has not been demonstrated.  If the true Poynting flux were
an order of magnitude smaller, so $\sim 10^{31}$ ergs day$^{-1}$, this
would represent a enough magnetic energy to power a large flare.  This
would contradict the view advocated by \citet{McClymont1989}, who
argued that even organized, near-surface flows could not inject enough
free magnetic energy into the coronal field to power large flares.
Instead, they suggested that the magnetic energy that is released in
very large flares must be carried into the corona with magnetic fields
as they emerge.

The view that flares are powered by free magnetic energy that emerges
with AR fields is, however, supported by the fact that active regions
which have recently flared are more likely flare again, compared to
active regions that have not recently flared. This phenomenon has been
referred to as persistence \citep{Zirin1991}.  The alternative view
requires post-emergence energy buildup in the corona, and is
plausible: flows could act on already emerged fields, leading to the
storage of free magnetic energy in the corona, which could then be
released in a flare.  But why would flows preferentially inject
magnetic energy into some ARs, making these flare more frequently, and
not into other ARs?  Indeed, it seems that some ARs emerge carrying a
lot free magnetic energy, which is released in many flares; such ARs
might be ``born bad'' (e.g., \citealt{Zirin1987}).  

But if flare-prone ARs are born bad, why is average flare flux
correlated with any photospheric flow property?  It could be that
$S_R$ is in some way related to the emergence of new, non-potential
magnetic fields.  But if emergence were truly the dominant manner of
free energy injection into the corona, then $\dot \Phi$ would be the
more fundamental magnetic field property, and therefore presumably a
better predictor of average flare flux.  (Perhaps our timescale for
computing $\dot \Phi$ --- nominally, 96 minutes --- is too short; see
below for more on this point.)  It could be that those ARs that emerge
with the largest amount of free magnetic energy, due to the presence
of electric currents, are also the most likely to exhibit large flows
--- perhaps as a result of Lorentz forces arising from these currents.

In addition, it is possible that cancellation, driven by seemingly
random flows, might lead to the buildup of free magnetic energy in the
corona, which might then be released in flares and CMEs.  This process
was not considered by \citet{McClymont1989}.  Turbulent convective
flows tend to systematically drive separate photospheric magnetic
features into close proximity in downflow lanes between supergranules
(e.g., \citealt{Schrijver1997}); while same-sign features simply
merge, opposite-sign features undergo cancellation.  Barring
unrealistic symmetry, such cancellation generally forces reconnection
between coronal fields, which has the effect of ``stranding'' any
magnetic helicity that was present in those fields in the corona, and
tends to increase the magnetic free energy of the coronal field
\citep{Welsch2006}.  The buildup of free magnetic energy by this
process, however, might be too slow to explain many large flares,
which can recur frequently in the same active region.

We were unable to clearly associate any intensive properties of flows
with the occurrence of flares.  Taken at face value, this might
suggest that the flare mechanism itself is fundamentally extensive, as
in the avalanche model of \citet{Lu1991}, for example.  We must,
however, qualify this result in several ways.  As discussed
above, our measure of flaring activity is extensive, which probably
biases our analysis against detecting relationships between intensive
flow properties and flares.  In addition, our flow estimates are
almost certainly subject to serious systematic errors.  Tests of FLCT,
DAVE, and other flow estimation techniques by \citet{Welsch2007} with
sequences of synthetic magnetograms from MHD simulations, in which the
true velocities are known, showed that both FLCT and DAVE
systematically underestimated the fluxes of magnetic helicity and
energy.  (These shortcomings led to the development of DAVE4VM
[\citealt{Schuck2008}], which uses information about the horizontal
magnetic field --- unavailable in this study --- to make improved
estimates of flows and the fluxes of magnetic energy and helicity.)
Moreover, these tests also showed that FLCT systematically
underestimated flow speeds.  So the tracking methods and/ or data (2''
spatial resolution, 96-min. cadence) employed in this study might be
too limited to reveal any intensive flow patterns that are associated
with flares.  Or it might be that intensive flow properties are
important, in some cases, for increasing coronal free magnetic energy,
and/ or for triggering flares, but that the efficacy of a
given flow pattern at injecting significant energy or triggering
flares depends upon the details of the particular photospheric
and coronal magnetic field configuration.  In this case, statistical
relationships between flows and flares would be hard to identify.  This last
possibility motivates development of time-dependent models of coronal
magnetic field evolution, driven by magnetograms, as tools necessary
to understand (and eventually predict) the coronal response to
photospheric flows.

Could it be that some flow patterns are, in fact, {\em
anti-correlated} with flare onset?  Some previous results have
suggested an inverse correlation between flow vorticity and flare
occurrence.  In his talk at the SHINE meeting in 2007, A.~A.~Pevtsov
presented preliminary results suggesting that flares are more likely
to occur {\em after} a time interval in which sunspots were observed
to rotate.  He presented a basic physical picture to explain these
observations: (1) the solar interior acts as a current driver that
imposes a fixed amount of twist (equivalently, current) on the flux
systems that comprise active regions; (2) if the twist in the coronal
portion of a flux system does not match the twist on the portion in
the interior, twist will propagate to equilibrate the two portions
\citep{Longcope2000}; (3) the twist propagation will be seen as a
rotation of photospheric footpoints, which will slow as the twist
gradient across the photosphere shrinks; (4) the twist injected into
the corona implies that free magnetic energy is present in the corona,
which will be released in a flare or flares.  Based upon this model,
one should expect that, statistically, flares will be more likely when
the corona has more twist, which should occur when footpoint rotation has
begun to slow or has ceased.

This preliminary investigation of the statistical relationships
between flows and flares has provided useful insights.  We have
investigated the properties of AR photospheric flows, characterized
the correlation times of AR fields and flows, and identified a
quantity, $S_R$, that might be useful for flare prediction.

Further study will be necessary to fully understand
the role of photospheric evolution in solar flares.  A study similar
to that conducted here, but with a larger sample of ARs, could verify
that our results are not statistical aberrations, due to our small
sample size.  Also, our tracking techniques are primarily senstive to
flows on the spatial scale of the windowing parameter used by DAVE and
FLCT, $\sim 17''$.  Flow-flare relationships on smaller and larger
spatial scales should also be investigated; perhaps particular flows
on very small scales, or coherent flows on large scales, are
important.  

In addition, evolution on other time scales should also be
investigated.  Our correlation analysis suggested that flux emergence,
quantified by the change in flux $\dot \Phi$ over our nominal
96-minute magnetogram cadence, was correlated with average flare flux,
but not as strongly as other field and flow properties that we
considered.  But fractional changes of flux in magnetograms are not
large on this time scale, which is short relative to the
characteristic hours-long emergence timescale of active region flux.
Consequently, the statistical relationship between flux emergence and
flares on longer time scales (e.g., 6 -- 48 hours) could reveal
stronger correlations with flare activity.  In addition, the
decorrelation rate of magnetograms (see Figs. \ref{fig:tcorr_flct} -
\ref{fig:tcorr_dave}) separated in time by intervals longer than our
nominal cadence could provide a quantitative measure of AR magnetic
evolution that might also be correlated with flares.  Work on these
efforts is underway.

\acknowledgments The authors appreciate the referee's careful reading
of the manuscript and thoughtful comments, which have improved the
manuscript.  B.~T. Welsch and Y. Li acknowledge the support of NSF
grant ATM-0451438, and NSF SHINE award ATM-0752597 for B.~T.~W.
Y.~L. additionally acknowledges support from NSF CISM ATM-0120950.
P.~W. Schuck acknowledges support from NASA LWS TR\&T grant
NNH06AD87I, LWS TR\&T Strategic Capability grant NNH07AG26I, and ONR.
G.~H. Fisher acknowledges support for his development work on FLCT
from NSF awards ATM-0641303 and ATM-0551084.  The discriminant
analysis code we used was kindly supplied by G. Barnes and K.~D. Leka,
who received funding from AFOSR under contracts F49620-00-C-0004,
F49620-03-C-0019, and from NASA under contract NNH07CD25C.  We thank
the SOHO/MDI team for making their database available and easy to use.
MDI is funded through NASA's Solar and Heliospheric Physics program.
The SOHO project results from international cooperation between NASA
and ESA.  B.~T.~W. also gratefully acknowledges discussions with
Dr. Maureen Lahiff regarding statistical analysis techniques.

\appendix

\section{Mercator Projections}
\label{app:merc}

A map projection is a two-dimensional, planar representation of a
two-dimensional, spherical surface.  In solar physics, instruments
typically record a two-dimensional image of the Sun in some wavelength
range as a plane-of-sky (POS) map of intensity variations.  When the
recorded emission originates in a relatively narrow atmospheric layer
--- as in a typical photospheric magnetogram, for instance --- this
procedure amounts to a projection from the imaged, spherical
atmospheric layer on the observer's side of the Sun onto the image
plane.  From the viewpoint of scientific analysis, this projection
poses many problems; for instance, foreshortening distorts spatial
relationships between image features.

To ameliorate this and other undesirable properties of the POS
projection, the POS positions of image data points can be converted to
latitude and longitude in some heliographic coordinate system.  The
origin of the Stonyhurst heliographic coordinate system (HCS), for
instance, is at the intersection of the apparent solar equator and the
central meridian as seen from Earth \citep{Thompson2006}.  This
conversion from POS to HCS coordinates is a deprojection.
Unfortunately, after transformation into longitude and latitude
coordinates $(\phi,\tilde{\theta})$ for a given heliographic
coordinate system, data that were regularly spaced in the POS
projection are typically not regularly sampled in
$(\phi,\tilde{\theta})$.  (We use $\tilde{\theta}$ for latitude, to
distinguish it from the co-latitude $\theta$ typically used in
spherical polar coordinates.)  Consequently, analyses of spatial
relationships between image features are still problematic.

Quantitative analyses of the spatial properties of irregularly sampled
data lying on a spherical surface therefore require interpolation of
these data onto a regular coordinate system.  While interpolation onto
spherical gridpoints uniformly spaced in $(\phi,\tilde{\theta})$ is
possible, both the distances and directions between nearby gridpoints
in such systems vary in complex ways in different regions of the grid.
Instead, we have opted to use the Mercator projection to convert the
data from $(\phi,\tilde{\theta})$ to a rectilinear coordinate system.
Often this two-projection procedure --- deprojecting from the POS
coordinates in the image plane to the HCS, and then transforming the
data again by interpolating onto spherical points corresponding to a
uniform grid --- is loosely referred to as ``deprojecting'' the data,
even though it is probably best described as a reprojection.  In the
discussion that follows, we assume the HCS positions of the POS image
points have already been computed, using, e.g., the IDL functions
\textsc{xy2lonlat.pro} in SSWIDL's /ssw/gen/idl/solar/~.

We chose the Mercator projection because it is conformal: locally, it
does not distort shape.  This means the Mercator projection should be
well-suited for unbiased (in direction) tracking of features in image
sequences, or any study of spatial variations in image structure,
e.g., the magnitude and direction of intensity gradients.  But, since
any projection of a spherical surface onto a plane must distort either
shape or scale (if not both!), the pixel scale varies over the
projected image.  (Equivalently, this projection is non-authalic, or
not area-preserving.)  We have developed our Mercator projection
procedure for tracking studies, on the assumption that correcting
distortion of scale is easier than correcting directional bias in
tracked displacements introduced by a non-conformal deprojection.

The distortion of scale in global Mercator projections is a function
of latitude alone.  The horizontal Mercator coordinate, $x$, is mapped
one-to-one with the heliocentric longitude, $\phi$, and is independent
of latitude.  Because the physical distance between lines of constant
longitude decreases as $\cos(\tilde \theta)$ with increasing latitude,
the projection's scale (the actual distance corresponding to a fixed
distance in the reprojected image) must decrease with increasing
latitude.  Displacements in the vertical Mercator coordinate, $dy$,
corresponding to a fixed latitudinal displacement, $d\tilde{\theta}$,
increase toward the poles.  Quantitatively, $y$ is related to latitude
$\tilde{\theta}$, via
\begin{equation}
y =  \ln \left ( \tan(\tilde{\theta}) + \sec(\tilde{\theta})\right ) ~.
\end{equation}
As $\tilde{\theta} \to \pi/2$, therefore, $y$ diverges
logarithmically.  Differentially, $dy = \sec(\tilde{\theta})
d\tilde{\theta}$, and the $dy$ corresponding to an interval
$d\tilde{\theta}$ also diverges toward the pole.  To compensate for
these distortions, small distances --- e.g., tracked displacements,
or the separations between pixel centers when computing gradients ---
must be multiplied by a factor of cosine latitude,
$\cos(\tilde{\theta})$, when converting Mercator distances to physical
distances.

It is desirable to preserve the spatial resolution in the original
data set.  Typically, this means the pixel width, $\Delta s$, in the
projected data should match the pixel size at the sub-observation
point (disk center) in the POS data.  Hence, this fixes the spatial
sampling rate of the original image that must be used to interpolate
the POS data onto the Mercator grid.  The number of necessary
longitudinal samples for an interval of longitude $\Delta \phi$ is the
horizontal distance corresponding to this interval, $L_\phi = R_\odot
\Delta \phi$, divided by the original pixel size, $\Delta s$.

To preserve spatial resolution in the latitidunal direction is more
complex. For a given interval of latitude, $\Delta \tilde{\theta} =
\tilde{\theta}_f - \tilde{\theta}_i$, the interval $\Delta y = y_f -
y_i$ is given by
\begin{equation}
\Delta y =  
\ln \left ( 
\frac{\tan(\tilde{\theta}_f) + \sec(\tilde{\theta}_f)}
     {\tan(\tilde{\theta}_i) + \sec(\tilde{\theta}_i)} \right ) ~.
\label{eqn:Deltay} \end{equation}
The number of spatial samples necessary to preserve the initial pixel
scale $\Delta s$ is therefore the distance corresponding to the
interval of latitude, $L_{\tilde{\theta}} = R_\odot \Delta
\tilde{\theta}$, divided by $\Delta s$, and multiplied by the ratio
$\Delta y/ \Delta \tilde{\theta}$.  For $ \Delta \tilde{\theta} \in
\{30^\circ, 45^\circ, 75^\circ\}$, the ratios $\Delta y/ \Delta
\tilde{\theta}$ are $\{1.05, 1.12, 2.02\}$, respectively.  Use of too
few samples in the vertical direction would distort shapes in the
image, and destroy the projection's conformal properties.

The necessary interpolation is done by first computing the Delaunay
triangulation of the irregularly sampled data in (latitude-longitude),
then interpolating onto the latitude-longitude points corresponding to
the regular Mercator grid of the output array.  (This output grid is
rectilinear in [latitude-longitude], but not regularly spaced in
latitude.)

When deprojecting sequences of image data, we found that small changes
in instrument pointing could lead to changes of +/- 1 pixel when
computing array sizes from pointing information in the header files.
Consequently, we fixed the output arrays for an entire image sequence,
as opposed to determining these separately for each image.

To verify that the algorithm worked properly, we projected latitudinal
and longitudinal bands onto a POS image of a sphere, and then
reprojected the POS image backward.

With the availability of full-disk vector magnetogram data from
SOLIS/VSM, SDO/HMI, and other instruments, conformal projections could
find wider application.  To view data near the poles without
distortion seen in typical global-scale Mercator projections (in which
the line of tangency between the sphere and projection cylinder lies
along the equator), local Mercator projections (in which the line of
tangency runs through the center of the deprojected region) could be
used, in a manner similar to the universal transverse Mercator (UTM)
approach used in terrestrial maps, e.g., NATO topographical maps.


\clearpage

\begin{deluxetable}{ccccc}
\small
\tablecaption{Active Region Sample}
\tablewidth{0pt}
\tablehead{
\colhead{ NOAA}  
& \colhead{Start Time} 
& \colhead{Stop Time}  
& \colhead{Avg. GOES Flux\tablenotemark{a}}
& \colhead{Avg. GOES Flux\tablenotemark{a}, $>$ C\tablenotemark{b}} \\
\colhead{ AR \#}  
& 
& 
& \colhead{($\mu$ W m$^{-2}$ day$^{-1}$)}
& \colhead{($\mu$ W m$^{-2}$ day$^{-1}$)}
}
\startdata
\hline
    7981 & 1996-08-01, 04:19 & 1996-08-04, 22:24 &  0.8 &  0.0 \\ 
    7986 & 1996-08-28, 04:51 & 1996-08-31, 22:24 &  0.4 &  0.0 \\ 
    7997 & 1996-11-22, 13:03 & 1996-11-23, 17:39 &  0.0 &  0.0 \\ 
    7999 & 1996-11-24, 06:27 & 1996-11-28, 20:48 &  8.7 &  8.0 \\ 
    8003 & 1996-12-08, 01:36 & 1996-12-11, 12:48 &  0.5 &  0.0 \\ 
    8004 & 1996-12-15, 08:03 & 1996-12-20, 06:27 &  0.1 &  0.0 \\ 
    8005 & 1996-12-16, 14:24 & 1996-12-21, 12:51 &  0.6 &  0.5 \\ 
    8006 & 1996-12-17, 08:00 & 1996-12-21, 15:59 &  0.0 &  0.0 \\ 
    8015 & 1997-01-30, 03:15 & 1997-02-03, 22:24 &  0.3 &  0.0 \\ 
    8020 & 1997-03-10, 20:48 & 1997-03-15, 03:12 &  0.0 &  0.0 \\ 
    8038 & 1997-05-09, 12:52 & 1997-05-14, 00:04 &  0.3 &  0.3 \\ 
    8040 & 1997-05-18, 08:04 & 1997-05-23, 12:52 &  3.1 &  2.9 \\ 
    8052 & 1997-06-13, 14:24 & 1997-06-17, 22:24 &  0.0 &  0.0 \\ 
    8076 & 1997-08-28, 03:11 & 1997-09-01, 09:39 &  2.5 &  1.9 \\ 
    8078 & 1997-08-29, 01:39 & 1997-09-01, 22:24 &  0.0 &  0.0 \\ 
    8081 & 1997-09-05, 01:39 & 1997-09-09, 12:48 &  0.0 &  0.0 \\ 
    8082 & 1997-09-05, 01:39 & 1997-09-09, 23:59 &  0.2 &  0.0 \\ 
    8083 & 1997-09-06, 14:23 & 1997-09-09, 12:48 &  3.7 &  3.3 \\ 
    8084 & 1997-09-08, 09:39 & 1997-09-13, 12:48 &  1.0 &  0.6 \\ 
    8085 & 1997-09-12, 01:36 & 1997-09-14, 06:24 &  1.4 &  1.2 \\ 
    8086 & 1997-09-15, 14:27 & 1997-09-20, 12:51 &  1.4 &  1.2 \\ 
    8088 & 1997-09-24, 08:03 & 1997-09-26, 22:24 &  45.4 &  44.2 \\ 
    8087 & 1997-09-26, 17:36 & 1997-09-29, 03:12 &  0.0 &  0.0 \\ 
    8090 & 1997-10-05, 11:15 & 1997-10-07, 06:27 &  0.0 &  0.0 \\ 
    8093 & 1997-10-10, 17:35 & 1997-10-15, 12:51 &  0.0 &  0.0 \\ 
    8096 & 1997-10-16, 17:35 & 1997-10-22, 08:03 &  0.0 &  0.0 \\ 
    8097 & 1997-10-19, 19:11 & 1997-10-24, 16:00 &  0.7 &  0.7 \\ 
    8100 & 1997-10-31, 01:39 & 1997-11-03, 23:59 &  88.6 &  87.2 \\ 
    8108 & 1997-11-17, 22:23 & 1997-11-22, 06:23 &  3.1 &  1.6 \\ 
    8123 & 1997-12-17, 01:39 & 1997-12-21, 22:17 &  0.1 &  0.0 \\ 
    8135 & 1998-01-11, 17:36 & 1998-01-15, 20:48 &  0.0 &  0.0 \\ 
    8156 & 1998-02-13, 17:36 & 1998-02-18, 22:24 &  3.9 &  2.8 \\ 
    8158 & 1998-02-15, 01:36 & 1998-02-18, 22:24 &  0.1 &  0.0 \\ 
    8179 & 1998-03-13, 03:12 & 1998-03-17, 22:24 &  20.6 &  19.7 \\ 
    8210 & 1998-04-28, 17:39 & 1998-05-03, 16:00 &  46.4 &  45.2 \\ 
    8214 & 1998-05-02, 17:36 & 1998-05-05, 16:00 &  11.2 &  10.7 \\ 
    8218 & 1998-05-10, 03:15 & 1998-05-14, 22:24 &  2.5 &  1.5 \\ 
    8222 & 1998-05-15, 17:36 & 1998-05-19, 03:15 &  0.5 &  0.5 \\ 
    8220 & 1998-05-15, 14:24 & 1998-05-19, 09:39 &  0.6 &  0.6 \\ 
    8227 & 1998-05-31, 01:39 & 1998-06-02, 12:51 &  0.0 &  0.0 \\ 
    8230 & 1998-06-02, 11:15 & 1998-06-06, 22:24 &  0.0 &  0.0 \\ 
    8243 & 1998-06-17, 01:36 & 1998-06-21, 12:48 &  4.6 &  4.1 \\ 
    8375 & 1998-11-02, 03:11 & 1998-11-06, 22:24 &  20.0 &  19.4 \\ 
    8377 & 1998-11-07, 03:12 & 1998-11-09, 19:15 &  0.0 &  0.0 \\ 
    8392 & 1998-11-26, 01:36 & 1998-11-29, 12:48 &  2.6 &  2.6 \\ 
    8393 & 1998-11-28, 01:36 & 1998-12-01, 22:24 &  0.4 &  0.4 \\ 
\enddata
\tablenotetext{a}{After \citet{Abramenko2005}}
\tablenotetext{b}{``$>$ C'' denotes only flare with peak flux $>$ GOES C1.0 level}
\label{tab:ar_data}
\normalsize 
\end{deluxetable}


\hoffset=-1.0in
\begin{deluxetable}{lcll}
\small
\tablecaption{Magnetic Field and Flow Properties Calculated}
\tablewidth{0pt}
\tablehead{
\colhead{ Property}  
& \colhead{Scaling\tablenotemark{a}}
& \colhead{Related Physical Quantity} 
& \colhead{Citation(s)\tablenotemark{b}} 
}
\startdata
\hline
$ \Phi $  & E  & Quantify AR ``magnetic size'' &  \citealt{Fisher1998};  \\ 
  &  &  & \citealt{Leka2003b} \\
$ \dot \Phi $  & E  & Flux emergence, cancellation &  \citealt{Priest1974};  \\ 
  &  &  & \citealt{Linker2003} \\
$ \dot \Phi / \Phi $  & I  & Flux emergence, cancellation &  \hspace{0.5in} ''  \\ 
$ |\dot \Phi|/ \Phi $  & I  & Unsigned flux changes &  \hspace{0.5in} ''  \\ 
$  {\cal M}[ \tilde{B}_R]  $ & I  & Field moments &  \citealt{Leka2003b}  \\ 
$  {\cal M}[ | \tilde{B}_R |]  $  & I  & Unsigned field moments &  \hspace{0.5in} '' \\
$  {\cal M}[ u]  $  & I  & Flow moments & ---   \\ 
$  {\cal M}[ u |d\Phi|/ \Phi_u]  $  & I  & Flux-weighted flow moments &  ---  \\ 
$ \sum u $  & E  & Amount of motion in AR & --- \\
$ \sum u|d\Phi| $  & E  & Unsigned flux-weighted speeds &  ---   \\ 
$  {\cal M}[|\hatr \cdot \nabla \times \uvec|]  $  & I  & Unsigned vorticity &  \citealt{Brown2003}  \\ 
$  {\cal M}[|d\Phi ( \hatr \cdot \nabla \times \uvec ) |/ \Phi_u]  $  & I  & Flux-weighted unsigned vorticity & \hspace{0.5in} '' \\
$ \sum \hatr \cdot (\nabla \times \uvec) $  & E  & Net vorticity &  \hspace{0.5in} ''  \\ 
$ \sum |\hatr \cdot (\nabla \times \uvec)| $  & E  & Unsigned vorticity &  \hspace{0.5in} ''  \\ 
$ \sum |d\Phi| ( \hatr \cdot \nabla \times \uvec ) $  &  E  & Flux-weighted vorticity &  \hspace{0.5in} ''  \\ 
$ \sum |d\Phi ( \hatr \cdot \nabla \times \uvec ) | $  & E  & Unsigned flux-weighted vorticity &  \hspace{0.5in} ''  \\ 
$  {\cal M}[ \nabla \cdot \uvec]  $  & I  & Converging/diverging flows  & \citealt{Linker2003}; \\
  &  &  & \citealt{Wang2006} \\
$  {\cal M}[|\nabla \cdot \uvec|]  $  & I  & Unsigned conv./div. flows  & \hspace{0.5in} '' \\
$  {\cal M}[|d\Phi| ( \nabla \cdot \uvec)/\Phi_u ]  $  & I  & Flux-weighted conv./div. flows  & \hspace{0.5in} '' \\
$  {\cal M}[|d\Phi (\nabla \cdot \uvec )|/\Phi_u]  $  & I  & Unsigned, flux-weighted conv./div. flows  &  \hspace{0.5in} '' \\
$ \sum (\nabla \cdot \uvec) $  & E  & conv./div. flows  & \hspace{0.5in} '' \\
$ \sum |\nabla \cdot \uvec| $  & E  & Unsigned conv./div. flows  & \hspace{0.5in} '' \\
$ \sum |d\Phi|(\nabla \cdot \uvec ) $  & E  & Flux-weighted, signed conv./div. flows  & \hspace{0.5in} '' \\
$ \sum |d\Phi (\nabla \cdot \uvec )| $  & E  & Unsigned, flux-weighted conv./div. flows  & \hspace{0.5in} '' \\
$ D_{\rm COF} $  & E  & Convergence of AR flux &  \citealt{Wang2006}  \\ 
$ S_{\rm COF} $  & E  & Shearing of AR flux &  \hspace{0.5in} '' \\
$ R $  & E  & Flux near PILs &  \citealt{Falconer2003}; \\
  &  &  & \citealt{Schrijver2007}  \\ 
$ \dot R $  & E  & Change in flux near PILs &  \hspace{0.5in} '' \\
$ \dot R / R $ & I  & Change in flux near PILs &  \hspace{0.5in} '' \\
$ S_R = \sum u \tilde{B}_R^2 $  & E  & Poynting flux proxy &  --- \\ 
$  {\cal M}[ u_c]  $  & I  & Flows along contours of $\tilde{B}_R$ &  \citealt{Demoulin2003}  \\ 
$  {\cal M}[ u_c|d\Phi|/ \Phi_u]  $  & I  & Flux-weighted contour flows &  \hspace{0.5in} '' \\
$  {\cal M}[ u_g]  $  & I  & Flows along gradients of $\tilde{B}_R$ &  --- \\ 
$  {\cal M}[ u_g|d\Phi|/ \Phi_u] $  & I  & Flux-weighted gradient flows &  --- \\ 
$  {\cal M}[ W u_g \tilde{B}_r]  $  & I  & Converging motions along PIL &  \citealt{Linker2003}  \\ 
$ \sum W u_g \tilde{B}_r  $  & E  & \hspace{0.5in} ''  &  \hspace{0.5in} ''  \\ 
$  {\cal M}[ W u_c \tilde{B}_r]  $  & I  & Shearing motions along PIL &  \citealt{Antiochos1999a}  \\ 
$ \sum W u_c \tilde{B}_r  $  & E  & \hspace{0.5in} '' & \hspace{0.5in} '' \\
$ \sum \tilde{B}_R^2  $  & E  & Component of Photospheric Magnetic Energy & ---  \\ 
$ \sum u^2 $  & E  & Kinetic energy & ---  \\ 
$ \sum | \tilde{B}_R | u^2 $  & E  & Field-weighted kinetic energy &  ---  \\ 
\enddata
\tablenotetext{a}{I $=$ intensive, E $=$ extensive}
\tablenotetext{b}{References chosen are representative, not necessarily either exhaustive or initial.}
\label{tab:props}
\normalsize 
\end{deluxetable}
\hoffset=0.25in



\begin{deluxetable}{cccccc}
\tablecaption{Magnetic Flux Correlations}
\tablewidth{0pt}
\tablehead{
& \colhead{$\Phi$}  
& \colhead{$R$} 
& \colhead{$\dot \Phi$}  
& \colhead{$\dot R$} 
& \colhead{$|\tilde{B}_R|^2$} 
}
\startdata
\hline
$\Phi$      
&   ---    
& $ 0.69$ \tablenotemark{b} 
& $ 0.18$ \tablenotemark{b} 
& $ 0.19$ \tablenotemark{b}  
& $ 0.91$ \tablenotemark{b} \\
 $R$         
& $ 0.53$ \tablenotemark{a}
&   ---    
& $ 0.60$ \tablenotemark{b} 
& $ 0.32$ \tablenotemark{b}  
& $ 0.84$ \tablenotemark{b} \\
 $\dot \Phi$ 
& $ 0.24$ \tablenotemark{a} 
& $ 0.54$ \tablenotemark{a} 
&   ---    
& $ 0.48$ \tablenotemark{b}  
& $ 0.40$ \tablenotemark{b} \\
 $\dot R$    
& $ 0.21$ \tablenotemark{a} 
& $ 0.54$ \tablenotemark{a} 
& $ 0.48$ \tablenotemark{a} 
&   ---     
& $ 0.17$ \tablenotemark{b} \\
 $|\tilde{B}_R|^2$ 
& $ 0.88$ \tablenotemark{a} 
& $ 0.61$ \tablenotemark{a} 
& $ 0.50$ \tablenotemark{a} 
& $ 0.23$ \tablenotemark{a}  
&   ---   
\enddata
\tablenotetext{a}{\, linear (Pearson) correlation}
\tablenotetext{b}{rank-order (Spearman) correlation}
\label{tab:flux_corrs}
\end{deluxetable}


\vspace{-2.5cm}
\begin{landscape}
\begin{deluxetable}{rclccccc}
\scriptsize
\tablecaption{Flare Frequencies and Skill Scores}
\tablewidth{0pt}
\tablehead{
  \colhead{Wind.}
& \colhead{Code}
& \colhead{Variable(s) Considered [Disc. Coeff.]}
& \colhead{PF/F\tablenotemark{a}}
& \colhead{PNF/F}
& \colhead{PF/NF}  
& \colhead{PNF/NF}  
& \colhead{$SS$}  
}
\startdata
\hline
6C  & DAVE  & $R$ 
            & 79 & 237 & 47 & 2345 & 0.16 \\
    &       & $R$ [1.24], $S_R$ [1.12]
            & 102 & 214 & 67 & 2325 & 0.19 \\
    &       & $R$ [1.19], $S_R$ [0.77], $\lan |\tilde{B}_R| \ran$ [0.52]
            & 106 & 210 & 70 & 2322 & 0.20 \\
\hline
6C  & FLCT  & $S_R$ 
            & 103 & 213 & 76 & 2316 & 0.16 \\
    &       & $S_R$ [1.21], $R$ [1.16]
            & 107 & 209 & 70 & 2322 & 0.20 \\
    &       & $S_R$ [1.64], $R$ [1.11], $\sum |\tilde{B}_R| u^2$ [-0.47]
            & 107 & 209 & 68 & 2324 & 0.20 \\
\hline
6N  & DAVE  & $R$ 
            & 39 & 80 & 62 & 2527 & -0.08 \\
    &       & $R$ [2.01], $S_R$ [0.76]
            & 40 & 79 & 72 & 2517 & -0.08 \\
    &       & $R$ [1.98], $S_R$ [0.60], $\lan |\tilde{B}_R| \ran$ [0.44]
            & 42 & 77 & 67 & 2522 & -0.05 \\
\hline
6N  & FLCT  & $R$ 
            & 39 & 80 & 62 & 2527 & -0.08 \\
    &       & $R$ [2.01], $S_R$ [0.71]
            & 41 & 78 & 66 & 2523 & -0.07 \\
    &       & $R$ [1.99], $S_R$ [0.53], $\Phi$ [0.47]
            & 42 & 77 & 64 & 2525 & -0.04 \\
\hline
24C  & DAVE  & $S_R$ 
            & 290 & 383 & 68 & 1967 & 0.33 \\
    &       & $S_R$ [1.19], $\lan |\tilde{B}_R| \ran$ [1.08]
            & 348 & 325 & 101 & 1934 & 0.37 \\
    &       & $S_R$ [1.63], $\lan |\tilde{B}_R| \ran$ [0.95], $\sum |\tilde{B}_R| u^2$ [-0.39]
            & 349 & 324 & 90 & 1945 & 0.37 \\
\hline
24C  & FLCT  & $S_R$ 
            & 299 & 374 & 72 & 1963 & 0.33 \\
    &       & $S_R$ [1.22], $\lan |\tilde{B}_R| \ran$ [1.03]
            & 335 & 338 & 76 & 1959 & 0.37 \\
    &       & $S_R$ [1.19], $\lan |\tilde{B}_R| \ran$ [0.71], $\sigma[\tilde{B}_R|]$ [0.37]
            & 330 & 343 & 76 & 1959 & 0.37 \\
\hline
24N  & DAVE  & $S_R$ 
            & 107 & 221 & 83 & 2297 & 0.15 \\
    &       & $S_R$ [1.19], $R$ [0.85]
            & 106 & 222 & 72 & 2308 & 0.17 \\
    &       & $S_R$ [0.82], $R$ [0.78], $\lan |\tilde{B}_R| \ran$ [0.55]
            & 105 & 223 & 73 & 2307 & 0.18 \\
\hline
24N  & FLCT  & $S_R$ 
            & 107 & 221 & 74 & 2306 & 0.16 \\
    &       & $S_R$ [1.34], $R$ [0.74]
            & 112 & 216 & 67 & 2313 & 0.18 \\
    &       & $S_R$ [1.35], $R$ [0.47], ${\cal F}_{\rm 24, current}$ [0.45] 
            & 115 & 213 & 63 & 2317 & 0.20 \\
\enddata 
\tablenotetext{a}{PF = predicted to flare; F = flare(s) occurred; %
PNF = predicted no flare; NF = no flare occurred}
\label{tab:skills}
\normalsize
\end{deluxetable}
\end{landscape}



\begin{landscape}
\begin{deluxetable}{llcccccc}
\tablecaption{Means of Selected Raw and Standardized Variables for Flaring and Non-Flaring Populations}
\tablewidth{0pt}
\tablehead{
  \colhead{Window}
&  \colhead{Variable [Units]\tablenotemark{a}}
& \colhead{$\overline{\rm F}$\tablenotemark{b}, raw}
& \colhead{$\overline{\rm NF}$\tablenotemark{c}, raw}
& \colhead{$\overline{\rm F+NF}$, raw}
& \colhead{Std. Dev.\tablenotemark{d}}
& \colhead{$\overline{\rm F}$, standardized\tablenotemark{e}}
& \colhead{$\overline{\rm NF}$, standardized}
}
\startdata
%
6C & $S_R$ [km G$^2$/s] &
3.6 $\times 10^{7}$ & 1.4 $\times 10^{7}$ & 1.6 $\times 10^{7}$ & 1.6 $\times 10^{7}$ & 1.3 & -1.7 $\times 10^{-1}$ \\
  & $R$ [G] &
7.3 $\times 10^{3}$ & 1.0 $\times 10^{3}$ & 1.8 $\times 10^{3}$ & 4.5 $\times 10^{3}$ & 1.2 & -1.6 $\times 10^{-1}$ \\
6N  & $R$ [G] &
1.0 $\times 10^{4}$ & 1.4 $\times 10^{3}$ & 1.8 $\times 10^{3}$ & 4.5 $\times 10^{3}$ & 1.9 & -8.8 $\times 10^{-2}$ \\
 & $S_R$ [km G$^2$/s] &
4.0 $\times 10^{7}$ & 1.5 $\times 10^{7}$ & 1.6 $\times 10^{7}$ & 1.6 $\times 10^{7}$ & 1.5 & -7.0 $\times 10^{-2}$ \\
24C & $S_R$  [km G$^2$/s] &
3.2 $\times 10^{7}$ & 1.1 $\times 10^{7}$ & 1.6 $\times 10^{7}$ & 1.6 $\times 10^{7}$ & 0.99  & -3.3 $\times 10^{-1}$ \\
 & $\lan \,|\tilde{B}_R|\, \ran$  [G] &
5.1 $\times 10^{1}$ & 3.1 $\times 10^{1}$ & 3.6 $\times 10^{1}$ & 1.6 $\times 10^{1}$ & 0.97  & -3.2 $\times 10^{-1}$ \\
24N & $S_R$  [km G$^2$/s] &
3.6 $\times 10^{7}$ & 1.4 $\times 10^{7}$ & 1.6 $\times 10^{7}$ & 1.6 $\times 10^{7}$ & 1.2 & -1.7 $\times 10^{-1}$ \\
 & $R$  [G] &
6.5 $\times 10^{3}$ & 1.1 $\times 10^{3}$ & 1.8 $\times 10^{3}$ & 4.5 $\times 10^{3}$ & 1.1 & -1.5 $\times 10^{-1}$ \\
%
\enddata 
\tablenotetext{a}{FLCT flow estimates were used.}
\tablenotetext{b}{Overlines denote averages; F = flare(s) occurred}
\tablenotetext{c}{NF = no flare occurred} 
\tablenotetext{d}{Standard deviation of combined F+NF populations.} 
\tablenotetext{e}{Derived by subtracting the mean from raw data, then 
dividing by the standard deviation.}
\label{tab:popstats}
\end{deluxetable}
\end{landscape}


\clearpage 
\eject

\clearpage 
\eject

\clearpage
\eject


\clearpage
\footnotesize
\normalsize
\eject


\clearpage
\eject


\clearpage
\eject

\clearpage
\eject


\clearpage
\eject

\clearpage
\eject


\clearpage
\eject

\clearpage
\eject

\clearpage
\eject

\clearpage
\eject


\footnotesize


\begin{thebibliography}{74}
\expandafter\ifx\csname natexlab\endcsname\relax\def\natexlab#1{#1}\fi

\bibitem[{{Abbett} \& {Fisher}(2003)}]{Abbett2003}
{Abbett}, W.~P. \& {Fisher}, G.~H. 2003, \apj, 582, 475

\bibitem[{Abbett {et~al.}(2000)Abbett, Fisher, \& Fan}]{Abbett2000}
Abbett, W.~P., Fisher, G.~H., \& Fan, Y. 2000, ApJ, 540, 548

\bibitem[{{Abramenko}(2005)}]{Abramenko2005}
{Abramenko}, V.~I. 2005, \apj, 629, 1141

\bibitem[{{Amari} {et~al.}(2003){Amari}, {Luciani}, {Aly}, {Mikic}, \&
  {Linker}}]{Amari2003a}
{Amari}, T., {Luciani}, J.~F., {Aly}, J.~J., {Mikic}, Z., \& {Linker}, J. 2003,
  ApJ, 585, 1073

\bibitem[{{Andrews}(2003)}]{Andrews2003}
{Andrews}, M.~D. 2003, \solphys, 218, 261

\bibitem[{{Antiochos} {et~al.}(1999){Antiochos}, {DeVore}, \&
  {Klimchuk}}]{Antiochos1999a}
{Antiochos}, S.~K., {DeVore}, C.~R., \& {Klimchuk}, J.~A. 1999, ApJ, 510, 485

\bibitem[{{Barnes} \& {Forecasting the Operational All-clear
  Participants}(2009)}]{Barnes2009}
{Barnes}, G. \& {Forecasting the Operational All-clear Participants}. 2009, in
  AAS/Solar Physics Division Meeting, Vol.~40, AAS/Solar Physics Division
  Meeting, \#16.05$--+$

\bibitem[{{Barnes} \& {Leka}(2008)}]{Barnes2008}
{Barnes}, G. \& {Leka}, K.~D. 2008, \apjl, 688, L107

\bibitem[{{Barnes} {et~al.}(2007){Barnes}, {Leka}, {Schumer}, \&
  {Della-Rose}}]{Barnes2007}
{Barnes}, G., {Leka}, K.~D., {Schumer}, E.~A., \& {Della-Rose}, D.~J. 2007,
  Space Weather, 5, 9002

\bibitem[{{Brown} {et~al.}(2003){Brown}, {Nightingale}, {Alexander},
  {Schrijver}, {Metcalf}, {Shine}, {Title}, \& {Wolfson}}]{Brown2003}
{Brown}, D.~S., {Nightingale}, R.~W., {Alexander}, D., {Schrijver}, C.~J.,
  {Metcalf}, T.~R., {Shine}, R.~A., {Title}, A.~M., \& {Wolfson}, C.~J. 2003,
  \solphys, 216, 79

\bibitem[{{Canfield} {et~al.}(1999){Canfield}, {Hudson}, \&
  {McKenzie}}]{Canfield1999}
{Canfield}, R.~C., {Hudson}, H.~S., \& {McKenzie}, D.~E. 1999, \grl, 26, 627

\bibitem[{{Chae} {et~al.}(2004){Chae}, {Moon}, \& {Park}}]{Chae2004b}
{Chae}, J., {Moon}, Y.-J., \& {Park}, Y.-D. 2004, \solphys, 223, 39

\bibitem[{Crosby {et~al.}(1993)Crosby, Ascwander, \& Dennis}]{Crosby1993}
Crosby, N.~B., Ascwander, M.~J., \& Dennis, B.~R. 1993, Solar~Phys., 143, 275

\bibitem[{{D{\' e}moulin} \& {Berger}(2003)}]{Demoulin2003}
{D{\' e}moulin}, P. \& {Berger}, M.~A. 2003, Sol.~Phys., 215, 203

\bibitem[{{D{\'e}moulin} \& {Pariat}(2009)}]{Demoulin2009}
{D{\'e}moulin}, P. \& {Pariat}, E. 2009, Advances in Space Research, 43, 1013

\bibitem[{{Deng} {et~al.}(2006){Deng}, {Xu}, {Yang}, {Cao}, {Liu}, {Rimmele},
  {Wang}, \& {Denker}}]{Deng2006}
{Deng}, N., {Xu}, Y., {Yang}, G., {Cao}, W., {Liu}, C., {Rimmele}, T.~R.,
  {Wang}, H., \& {Denker}, C. 2006, \apj, 644, 1278

\bibitem[{{Falconer} {et~al.}(2003){Falconer}, {Moore}, \&
  {Gary}}]{Falconer2003}
{Falconer}, D.~A., {Moore}, R.~L., \& {Gary}, G.~A. 2003, Journal of
  Geophysical Research (Space Physics), 108, 11

\bibitem[{{Falconer} {et~al.}(2006){Falconer}, {Moore}, \&
  {Gary}}]{Falconer2006}
---. 2006, \apj, 644, 1258

\bibitem[{{Fan}(2001)}]{Fan2001}
{Fan}, Y. 2001, \apjl, 554, L111

\bibitem[{{Fan} {et~al.}(2003){Fan}, {Abbett}, \& {Fisher}}]{Fan2003}
{Fan}, Y., {Abbett}, W.~P., \& {Fisher}, G.~H. 2003, \apj, 582, 1206

\bibitem[{Fisher {et~al.}(1998)Fisher, Longcope, Metcalf, \&
  Pevtsov}]{Fisher1998}
Fisher, G.~H., Longcope, D.~W., Metcalf, T.~R., \& Pevtsov, A.~A. 1998, ApJ,
  508, 885

\bibitem[{{Fisher} \& {Welsch}(2008)}]{Fisher2008}
{Fisher}, G.~H. \& {Welsch}, B.~T. 2008, in Astronomical Society of the Pacific
  Conference Series, Vol. 383, Astronomical Society of the Pacific Conference
  Series, ed. R.~{Howe}, R.~W. {Komm}, K.~S. {Balasubramaniam}, \& G.~J.~D.
  {Petrie}, 373--380; also arXiv:0712.4289

\bibitem[{{Forbes}(2000)}]{Forbes2000}
{Forbes}, T.~G. 2000, JGR, 105, 23153

\bibitem[{{Harvey} \& {Zwaan}(1993)}]{Harvey1993}
{Harvey}, K.~L. \& {Zwaan}, C. 1993, \solphys, 148, 85

\bibitem[{{Hathaway} {et~al.}(2002){Hathaway}, {Beck}, {Han}, \&
  {Raymond}}]{Hathaway2002}
{Hathaway}, D.~H., {Beck}, J.~G., {Han}, S., \& {Raymond}, J. 2002, \solphys,
  205, 25

\bibitem[{Hills(1966)}]{Hills1966}
Hills, M. 1966, Journal of the Royal Statistical Society, Series B
  (Methodological), 28, 1

\bibitem[{{Huberty} \& {Wisenbaker}(1992)}]{Huberty1992}
{Huberty}, C.~J. \& {Wisenbaker}, J.~M. 1992, Journal of Educational
  Statistics, 17, 75

\bibitem[{{Hudson}(1991)}]{Hudson1991}
{Hudson}, H.~S. 1991, \solphys, 133, 357

\bibitem[{{Keil} {et~al.}(1994){Keil}, {Balasubramaniam}, {Bernasconi},
  {Smaldone}, \& {Cauzzi}}]{Keil1994}
{Keil}, S.~L., {Balasubramaniam}, K.~S., {Bernasconi}, P., {Smaldone}, L.~A.,
  \& {Cauzzi}, G. 1994, in Astronomical Society of the Pacific Conference
  Series, Vol.~68, Solar Active Region Evolution: Comparing Models with
  Observations, ed. K.~S. {Balasubramaniam} \& G.~W. {Simon}, 265--+

\bibitem[{Knoll {et~al.}(2008)Knoll, Leka, \& Barnes}]{Knoll2008}
Knoll, J., Leka, K.~D., \& Barnes, G. 2008, LASP REU Student Presentations,
  unpublished,
  http://lasp.colorado.edu/reu/student\_docs/posters/Knoll\_Poster.pdf

\bibitem[{{Kucera} {et~al.}(1997){Kucera}, {Dennis}, {Schwartz}, \&
  {Shaw}}]{Kucera1997}
{Kucera}, T.~A., {Dennis}, B.~R., {Schwartz}, R.~A., \& {Shaw}, D. 1997, \apj,
  475, 338

\bibitem[{{LaBonte} {et~al.}(2007){LaBonte}, {Georgoulis}, \&
  {Rust}}]{LaBonte2007}
{LaBonte}, B.~J., {Georgoulis}, M.~K., \& {Rust}, D.~M. 2007, \apj, 671, 955

\bibitem[{Leighton {et~al.}(1962)Leighton, Noyes, \& Simon}]{Leighton1962}
Leighton, R.~B., Noyes, R.~W., \& Simon, G.~W. 1962, ApJ, 135, 474

\bibitem[{{Leka} \& {Barnes}(2003)}]{Leka2003b}
{Leka}, K.~D. \& {Barnes}, G. 2003, \apj, 595, 1296

\bibitem[{{Leka} \& {Barnes}(2007)}]{Leka2007}
---. 2007, \apj, 656, 1173

\bibitem[{Leka {et~al.}(1996)Leka, Canfield, McClymont, \& Van~{D}riel
  {G}esztelyi}]{Leka1996}
Leka, K.~D., Canfield, R.~C., McClymont, A.~N., \& Van~{D}riel {G}esztelyi, L.
  1996, ApJ, 462, 547

\bibitem[{{Li} {et~al.}(2004){Li}, {Luhmann}, {Fisher}, \& {Welsch}}]{Li2004}
{Li}, Y., {Luhmann}, J., {Fisher}, G., \& {Welsch}, B. 2004, Journal of
  Atmospheric and Solar-Terrestrial Physics, 66, 1271

\bibitem[{{Li} {et~al.}(2009){Li}, {Lynch}, {Welsch}, {Stenborg}, {Fisher},
  {Liu}, \& {Nightingale}}]{Li2009}
{Li}, Y., {Lynch}, B.~J., {Welsch}, B., {Stenborg}, G.~A.and~{Luhmann}, J.~G.,
  {Fisher}, G.~H., {Liu}, Y., \& {Nightingale}, R.~W. 2009, \solphys,
  submitted, "

\bibitem[{{Linker} {et~al.}(2001){Linker}, {Lionello}, {Miki{\' c}}, \&
  {Amari}}]{Linker2001}
{Linker}, J.~A., {Lionello}, R., {Miki{\' c}}, Z., \& {Amari}, T. 2001, JGR,
  106, 25165

\bibitem[{{Linker} {et~al.}(2003){Linker}, {Miki{\'c}}, {Lionello}, {Riley},
  {Amari}, \& {Odstrcil}}]{Linker2003}
{Linker}, J.~A., {Miki{\'c}}, Z., {Lionello}, R., {Riley}, P., {Amari}, T., \&
  {Odstrcil}, D. 2003, Physics of Plasmas, 10, 1971

\bibitem[{{Lionello} {et~al.}(2002){Lionello}, {Miki{\' c}}, {Linker}, \&
  {Amari}}]{Lionello2002}
{Lionello}, R., {Miki{\' c}}, Z., {Linker}, J.~A., \& {Amari}, T. 2002, ApJ,
  581, 718

\bibitem[{Longcope {et~al.}(2005)Longcope, McKenzie, Cirtain, \&
  Scott}]{Longcope2005}
Longcope, D.~W., McKenzie, D., Cirtain, J., \& Scott, J. 2005, \apj, 630, 596

\bibitem[{{Longcope} \& {Welsch}(2000)}]{Longcope2000}
{Longcope}, D.~W. \& {Welsch}, B.~T. 2000, ApJ, 545, 1089

\bibitem[{Lu \& Hamilton(1991)}]{Lu1991}
Lu, E.~T. \& Hamilton, R.~J. 1991, ApJ, 380, L89

\bibitem[{{Lynch} {et~al.}(2008){Lynch}, {Antiochos}, {DeVore}, {Luhmann}, \&
  {Zurbuchen}}]{Lynch2008}
{Lynch}, B.~J., {Antiochos}, S.~K., {DeVore}, C.~R., {Luhmann}, J.~G., \&
  {Zurbuchen}, T.~H. 2008, \apj, 683, 1192

\bibitem[{{Magara} \& {Longcope}(2001)}]{Magara2001}
{Magara}, T. \& {Longcope}, D.~W. 2001, ApJ, 559, L55

\bibitem[{{Manchester}(2007)}]{Manchester2007}
{Manchester}, W.~I. 2007, \apj, 666, 532

\bibitem[{Mc{C}lymont \& Fisher(1989)}]{McClymont1989}
Mc{C}lymont, A.~N. \& Fisher, G.~H. 1989, in Solar System Plasma Physics, ed.
  J.~H. Waite, J.~L. Burch, \& R.~L. Moore, AGU, Washington, 219

\bibitem[{{Moore} {et~al.}(2001){Moore}, {Sterling}, {Hudson}, \&
  {Lemen}}]{Moore2001}
{Moore}, R.~L., {Sterling}, A.~C., {Hudson}, H.~S., \& {Lemen}, J.~R. 2001,
  ApJ, 552, 833

\bibitem[{November \& Simon(1988)}]{November1988}
November, L. \& Simon, G. 1988, ApJ, 333, 427

\bibitem[{{Parker}(1984)}]{Parker1984a}
{Parker}, E.~N. 1984, \apj, 280, 423

\bibitem[{{Pevtsov} {et~al.}(2003){Pevtsov}, {Fisher}, {Acton}, {Longcope},
  {Johns-Krull}, {Kankelborg}, \& {Metcalf}}]{Pevtsov2003}
{Pevtsov}, A.~A., {Fisher}, G.~H., {Acton}, L.~W., {Longcope}, D.~W.,
  {Johns-Krull}, C.~M., {Kankelborg}, C.~C., \& {Metcalf}, T.~R. 2003, ApJ,
  598, 1387

\bibitem[{{Priest} \& {Heyvaerts}(1974)}]{Priest1974}
{Priest}, E.~R. \& {Heyvaerts}, J. 1974, \solphys, 36, 433

\bibitem[{{Roussev} {et~al.}(2004){Roussev}, {Sokolov}, {Forbes}, {Gombosi},
  {Lee}, \& {Sakai}}]{Roussev2004}
{Roussev}, I.~I., {Sokolov}, I.~V., {Forbes}, T.~G., {Gombosi}, T.~I., {Lee},
  M.~A., \& {Sakai}, J.~I. 2004, ApJ, 605, L73

\bibitem[{{Sammis} {et~al.}(2000){Sammis}, {Tang}, \& {Zirin}}]{Sammis2000}
{Sammis}, I., {Tang}, F., \& {Zirin}, H. 2000, \apj, 540, 583

\bibitem[{Scherrer {et~al.}(1995)Scherrer, Bogart, Bush, Hoeksema, Kosovichev,
  Schou, Rosenberg, Springer, Tarbell, Title, Wolfson, Zayer, \& {The MDI
  Engineering Team}}]{Scherrer1995}
Scherrer, P., Bogart, R.~S., Bush, R.~I., Hoeksema, J.~T., Kosovichev, A.,
  Schou, J., Rosenberg, W., Springer, L., Tarbell, T., Title, A., Wolfson, C.,
  Zayer, I., \& {The MDI Engineering Team}. 1995, Solar~Phys., 162, 129

\bibitem[{{Schrijver}(2007)}]{Schrijver2007}
{Schrijver}, C.~J. 2007, \apjl, 655, L117

\bibitem[{{Schrijver} {et~al.}(2005){Schrijver}, {DeRosa}, {Title}, \&
  {Metcalf}}]{Schrijver2005}
{Schrijver}, C.~J., {DeRosa}, M.~L., {Title}, A.~M., \& {Metcalf}, T.~R. 2005,
  \apj, 628, 501

\bibitem[{Schrijver {et~al.}(1997)Schrijver, Title, van Ballegooijen, Hagenaar,
  , \& Shine}]{Schrijver1997}
Schrijver, C.~J., Title, A.~M., van Ballegooijen, A.~A., Hagenaar, H.~J., , \&
  Shine, R.~A. 1997, ApJ, 487, 424

\bibitem[{{Schuck}(2005)}]{Schuck2005}
{Schuck}, P.~W. 2005, ApJ, 632, L53

\bibitem[{{Schuck}(2006)}]{Schuck2006}
---. 2006, \apj, 646, 1358

\bibitem[{{Schuck}(2008)}]{Schuck2008}
---. 2008, \apj, 683, 1134

\bibitem[{{Thompson}(2006)}]{Thompson2006}
{Thompson}, W.~T. 2006, \aap, 449, 791

\bibitem[{{Tokman} \& {Bellan}(2002)}]{Tokman2002}
{Tokman}, M. \& {Bellan}, P.~M. 2002, \apj, 567, 1202

\bibitem[{{Wang}(2006)}]{Wang2006}
{Wang}, H. 2006, \apj, 649, 490

\bibitem[{Welsch(2006)}]{Welsch2006}
Welsch, B.~T. 2006, ApJ, 638, 1101

\bibitem[{Welsch {et~al.}(2007)Welsch, Abbett, DeRosa, Fisher, Georgoulis,
  Longcope, Ravindra, \& Schuck}]{Welsch2007}
Welsch, B.~T., Abbett, W.~P., DeRosa, M.~L., Fisher, G.~H., Georgoulis,
  M.~K.~Kusano, K., Longcope, D.~W., Ravindra, B., \& Schuck, P.~W. 2007, ApJ,
  670, 1434

\bibitem[{{Welsch} \& {Fisher}(2008)}]{Welsch2008a}
{Welsch}, B.~T. \& {Fisher}, G.~H. 2008, in Astronomical Society of the Pacific
  Conference Series, Vol. 383, Astronomical Society of the Pacific Conference
  Series, ed. R.~{Howe}, R.~W. {Komm}, K.~S. {Balasubramaniam}, \& G.~J.~D.
  {Petrie}, 19--30; also arXiv:0710.0546

\bibitem[{Welsch {et~al.}(2004)Welsch, Fisher, Abbett, \&
  R\'egnier}]{Welsch2004}
Welsch, B.~T., Fisher, G.~H., Abbett, W.~P., \& R\'egnier, S. 2004, ApJ, 610,
  1148

\bibitem[{{Welsch} \& {Li}(2008)}]{Welsch2008b}
{Welsch}, B.~T. \& {Li}, Y. 2008, in Astronomical Society of the Pacific
  Conference Series, Vol. 383, Astronomical Society of the Pacific Conference
  Series, ed. R.~{Howe}, R.~W. {Komm}, K.~S. {Balasubramaniam}, \& G.~J.~D.
  {Petrie}, 429--437; also arXiv:0710.0562

\bibitem[{{Wheatland}(2001)}]{Wheatland2001}
{Wheatland}, M.~S. 2001, \solphys, 203, 87

\bibitem[{{Wheatland}(2005)}]{Wheatland2005}
---. 2005, Space Weather, 3, 7003

\bibitem[{{Zirin} \& {Liggett}(1987)}]{Zirin1987}
{Zirin}, H. \& {Liggett}, M.~A. 1987, \solphys, 113, 267

\bibitem[{{Zirin} \& {Marquette}(1991)}]{Zirin1991}
{Zirin}, H. \& {Marquette}, W. 1991, \solphys, 131, 149

\end{thebibliography}

\normalsize

\end{document}